\documentclass[10pt]{iopart}
\usepackage{iopams}  
\usepackage{graphicx}
\usepackage{color}
\usepackage{hyperref}
\usepackage{comment}
\usepackage[normalem]{ulem}
\newcommand{\pd}{\partial}
\newcommand{\bd}[1]{\boldsymbol{#1}}

\newcommand{\ie}{{i.e., }}
\newcommand{\mc}[1]{\mathcal{#1}}

\newcommand{\avg}[1]{\left\langle{#1}\right\rangle}

\newcommand{\ExB}{{\bd{E}\times \bd{B}}}
\newcommand{\rml}{{\rm l}}
\newcommand{\rmm}{{\rm m}}
\newcommand{\rmp}{{\rm p}}
\newcommand{\rmr}{{\rm r}}
\newcommand{\rmt}{{\rm t}}

\begin{document}

\title[]{Global linear drift-wave eigenmode structures on flux surfaces in stellarators: ion temperature gradient mode}

\author{Hongxuan Zhu$^1$, H. Chen$^2$, Z. Lin$^2$, and A. Bhattacharjee$^1$}
\address{$^1$Department of Astrophysical Sciences, Princeton University, Princeton, NJ 08544}
\address{$^2$Department of Physics and Astronomy, University of California, Irvine, CA 92697}
\begin{abstract}
Turbulent transport greatly impacts the performance of stellarator magnetic confinement devices. While significant progress has been made on the numerical front, theoretical understanding of turbulence in stellarators is still lacking. In particular, due to nonaxisymmetry, different field lines couple within flux surfaces, the effects from which have yet to be adequately studied. In this work, we numerically simulate the linear electrostatic ion-temperature-gradient modes in stellarators using the global gyrokinetic particle-in-cell code GTC. We find that the linear eigenmode structures are nonuniform across field lines on flux surfaces and are localized  at the downstream of the ion diamagnetic drift. Based on a simple model from Zocco \etal [Phys. Plasmas \textbf{23}, 082516 (2016); \textbf{27}, 022507 (2020)], we show that the localization can be explained from the nonzero imaginary part of the binormal wavenumber. We further demonstrate that a localized surface-global eigenmode can be constructed from local gyrokinetic codes \texttt{stella} and GX, only if we first solve the local dispersion relation with real wavenumbers on each field line, and then do an analytic continuation  to the complex-wavenumber plane. These results suggest that the complex-wavenumber spectra from surface-global effects are required to understand the linear drift-wave eigenmode structures in stellarators.
\end{abstract}
\maketitle
\ioptwocol
\section{Introduction}
\label{sec:introduction}
Turbulent transport significantly impacts the performance of stellarator magnetic confinement devices. For example, in the Wendelstein 7-X (W7-X) device, the ion-temperature-gradient (ITG) turbulence is believed to limit the achievable core ion temperature in electron-cyclotron-resonance-heated plasmas \cite{Beurskens21}. It is well known that plasma microturbulence is highly anisotropic in magnetic confinement devices, $l_\parallel\gg l_\perp$, where $l_\parallel$ ($l_\perp$) is the characteristic wavelength along (across) the magnetic fields $\bd{B}$. Therefore, fluctuating quantities such as the electrostatic potential $\Phi$ can be written as $\hat{\Phi}(\psi,\alpha,l)\rme^{\rmi S(\psi,\alpha)/\rho_*}$, which consists of a rapidly varying phase factor $\rme^{\rmi S/\rho_*}$ and a slowly varying envelope $\hat{\Phi}$. Here, $\psi$ is the flux-surface label, $\alpha$ is the field-line label, $l$ is the distance along field lines,  and $\rho_*=\rho_\rmi/a$ is a small parameter with $\rho_\rmi$ the ion gyroradius at thermal velocity and $a$ the minor radius of the device.  This motivates the local (flux-tube) approach, which writes $\nabla S=k_\psi\nabla\psi+k_\alpha\nabla\alpha$  with real radial real wavenumber $k_\psi$ and binormal real wavenumber $k_\alpha$, so that one can focus on the structure of $\hat{\Phi}$ as a function of $l$, while treating $\psi$ and $\alpha$ as parameters \cite{Beer95,Highcock12}. In other words, the local approach studies fluctuations in a slender parallelogram tube along a given field line (so-called flux tube) at given $(\psi,\alpha)$ and ignores effects from coupling among different field lines. Proper boundary conditions are required to treat the two ends of the flux tube, and the twist-and-shift boundary condition has been widely used for tokamak simulations \cite{Beer95}.  For  stellarator simulations, the twist-and-shift boundary condition could fail due to the low global magnetic shear, and the generalized twist-and-shift boundary condition \cite{Martin18} has often been used instead.

Stellarators are nonaxisymmeric and the turbulent fluctuation level generally depends on both $l$ and $\alpha$.  (We focus on the fluctuation level within flux surfaces, so that the dependence on $\psi$ will not be discussed in this paper.)  Such dependence could come from either the local effects, where the local geometric quantities vary with $\alpha$, or the global effects, where fluctuations at different $\alpha$ couple together. To fully exploit the local effects, several flux tubes at different $\alpha$, or a single flux tube with several poloidal turns, are often required for local simulations. The choice of $\alpha$, however, can be restrictive. For example, the generalized boundary condition, which requires that fluctuations at the two ends of the flux tube behave statistically the same, only applies to flux tubes that possess stellarator symmetry \cite{Dewar98}. For this reason, local simulations often choose the $\alpha=0$ and the $\alpha=\iota\pi/N_{\rm fp}$ flux tubes that possess stellarator symmetry, and extend the length of them to many poloidal turns, so that the flux tubes amply sample the geometric information on the flux surfaces \cite{Smoniewski21,Sanchez21} ($\iota$ is the rotational transform and $N_{\rm fp}$ is the field period of the device).  In other words, one essentially simulates the whole flux surface, except that turbulent correlation only happens in the parallel direction.

Despite the efforts in predicting turbulent transport from local simulations, there are open questions regarding the observed discrepancy between local and global simulation results on the turbulent fluctuation level in stellarators  \cite{Xanthopoulos14,Wilms23,Wilms24}. In fact, global gyrokinetic simulations  found that the linear ITG eigenmodes are highly localized in $\alpha$, which is a common phenomenon in quasi-isodynamic (QI)  W7-X configurations \cite{Kornilov04,Sanchez19,Cole19,Wang20,BanonNavarro20,Sanchez23}, as well as in quasi-axisymmetric (QA) and quasi-helically symmetric (QH) configurations \cite{Jost01,Cole20,Chen25a}. Recent studies also found similar localization in $\alpha$ for the linear trapped-electron mode in QH  \cite{Chen25b} and QI \cite{Nicolau25} configurations. Since local simulations determine the linear eigenmode structures in $l$ but not in $\alpha$, the localization of linear ITG eigenmodes in $\alpha$ must be a global effect. For the nonlinear stage of the ITG turbulence, it remains unclear whether local  or global effects are more important. For example, full-flux-surface GENE simulations of ITG turbulence in a QA configuration found that the localization in $\alpha$ persists in the nonlinear stage at a smaller $\rho_*=1/250$ but not at a larger $\rho_*=1/125$ \cite{Xanthopoulos16}. A heuristic explanation of the difference between local and global simulations has been given in \cite{Helander15}. More recently, from global GENE-3D simulations of ITG turbulence in W7-X \cite{BanonNavarro20,Sanchez23}, the fluctuation level noticeably deviates from stellarator symmetry, which local simulation results would always obey. Besides, turbulence can spread nonlinearly in both the poloidal and the radial directions \cite{Chen25a,Nicolau25}. Therefore, a careful study on the global effects is still needed.

In this work, we numerically simulate the linear electrostatic ITG eigenmodes in stellarators using the global gyrokinetic particle-in-cell code GTC, and present a theoretical explanation for the observed mode structures. We simulate the precise QA and precise QH configurations reported in \cite{Landreman22}, as well as a W7-X high-mirror configuration used in \cite{Gonzalez22}.  We find  that the linear eigenmode structures are nonuniform in $\alpha$ on flux surfaces and are localized  at the downstream of the ion diamagnetic drift. Based on a simple model from Zocco \etal \cite{Zocco16,Zocco20} and following the WKB theory of Dewar and Glasser \cite{Dewar83}, we show that the localization can be explained from the nonzero imaginary part of $k_\alpha$. Focusing on the precise QA configuration, we further demonstrate that a localized surface-global eigenmode can be constructed from local gyrokinetic codes \texttt{stella} \cite{Barnes19} and GX \cite{Mandell24}, only if we first solve the local dispersion relation with real wavenumbers, and then do an analytic continuation  to the complex-wavenumber plane. These results suggest that the complex-wavenumber spectra from surface-global effects are required to understand the linear drift-wave eigenmode structures in stellarators (while existing local simulations always assume real wavenumbers).

In the following, global GTC simulation results of the linear ITG eigenmodes are given in \sref{sec:GTC}. A theoretical explanation of the mode localization based on  \cite{Zocco16,Zocco20} is given in \sref{sec:theory}. Comparisons between global and local simulation results are given in \sref{sec:local}. And conclusions are given in \sref{sec:conclusion}. A systematic derivation of the  mode structures across field lines from collisionless electrostatic gyrokinetic equations is given in \ref{sec:app}.
\section{GTC simulation of global linear ITG eigenmode structures}
\label{sec:GTC}
We use the global gyrokinetic particle-in-cell code GTC\footnote{\url{https://sun.ps.uci.edu/gtc}} to simulate linear electrostatic ITG eigenmodes in stellarators. The code utilizes Boozer coordinates, which are suitable for the nonaxisymmetric stellarator simulations \cite{Wang20,Fu21,Nicolau21,Singh23,Zhu25}. In particular, the code defines the poloidal and toroidal angle $(\theta,\zeta)$ so that the poloidal magnetic flux $\psi_\rmp$ is always positive. However, to better compare with local simulations, we will use $\psi=-\psi_{\rm V}/2\pi$ as the flux-surface label in the following, where $\psi_{\rm V}$ is the toroidal flux  from the VMEC equilibria\footnote{\url{https://princetonuniversity.github.io/STELLOPT/VMEC.html}}. For our simulations, $\psi$ is negative and decreases radially, $\theta$  increases counter-clockwise, and $\zeta$ is in the same direction as the cylindrical toroidal angle. Then, magnetic fields can be represented as
\begin{equation}
\bd{B}=\nabla\psi\times\nabla\alpha=\nabla\psi\times\nabla\theta+\iota\nabla\zeta\times\nabla\psi,
\end{equation}
where $\alpha=\theta-\iota\zeta$ is the field-line label and $\iota(\psi)$ is the rotational transform. For all three configurations the toroidal magnetic fields point in the positive-$\zeta$ direction, so that ion diamagnetic drift, which is proportional to $\bd{B}\times\nabla (n_\rmi T_\rmi)$, is counter-clockwise. The ion  drift frequency, which is proportional to $\bd{B}\times\nabla B\cdot \nabla\alpha$, is also positive in the so-called bad-curvature region where ITG modes are expected to reside.

We simulate gyrokinetic deuterium ions with mass $m_\rmi=2m_\rmp$ ($m_\rmp$ is the proton mass) and charge number $Z_\rmi=1$. Using spatial coordinates $\bd{R}$, magnetic moment $\mu=m_\rmi v_\perp^2/2B$ ($v_\perp$ is the perpendicular velocity), and energy $\mc{E}=\mu B+m_\rmi v_\parallel^2/2$ ($v_\parallel$ is the parallel velocity) as independent phase-space coordinates, the linearized delta-$f$ collisionless gyrokinetic equation for ions is
\begin{equation}
\label{eq:GTC_linear}
\eqalign{
    \frac{\pd\delta\!f}{\pd t}+(v_\parallel\hat{\bd{b}}&+\bd{v}_\rmd)\cdot\nabla\delta\!f+\bd{v}_E\cdot\nabla f_0\\
    &-Z_\rmi e\frac{\pd f_0}{\pd\mc{E}}(v_\parallel\hat{\bd{b}}+\bd{v}_\rmd)\cdot \nabla \delta\bar{\Phi}=0.
    }
\end{equation}
Here, the gyrocenter ion distribution is $f_\rmi=f_0+\delta\! f$ where $f_0$ is the equilibrium and $\delta\! f$ is the perturbation, $\delta\Phi$ is the fluctuating electrostatic potential, $\hat{\bd{b}}=\bd{B}/B$, $e$ is the elementary charge, $\bd{v}_\rmd=\Omega_\rmi^{-1}\hat{\bd{b}}\times(v_\perp^2\nabla\ln B/2+v_\parallel^2\hat{\bd{b}}\cdot\nabla\hat{\bd{b}})$ is the magnetic (grad-$B$ and curvature) drift with $\Omega_\rmi=eB/m_\rmi$, and $\bd{v}_E=\hat{\bd{b}}\times\nabla\delta\bar{\Phi}/B$ is the $\ExB$ drift from the gyroaveraged potential $\delta\bar{\Phi}$. Writing $\delta\Phi$ as the Fourier series, $\delta\Phi=\sum_{\bd{k}}\delta\Phi_{\bd{k}}\rme^{\rmi\bd{k}\cdot\bd{R}/\rho_*}$, the gyroaveraged potential can be expressed as
\begin{equation}
\delta\bar{\Phi}=\sum_{\bd{k}}\delta\Phi_{\bd{k}}J_0\left(\frac{k_\perp v_\perp}{\rho_*\Omega_\rmi}\right)\rme^{\rmi\bd{k}\cdot\bd{R}/\rho_*},
\end{equation}
where $J_0$ is the Bessel function of the first kind and $k_\perp$ is the perpendicular wavenumber. For our simulations, 
\begin{equation}
f_0=\frac{n_\rmi}{\pi^{3/2}v_{\rmt\rmi}^3}\rme^{-\mc{E}/T_\rmi}    
\end{equation} 
is a local Maxwellian, where the equilibrium density $n_\rmi(\psi)$ and temperature $T_\rmi(\psi)$ are flux functions, and the thermal velocity is $v_{\rmt\rmi}=\sqrt{2T_\rmi/m_\rmi}$. Electrons are assumed adiabatic, $\delta n_\rme/n_\rme=e\delta\Phi/T_\rme$, where $n_\rme=n_\rmi$ due to quasineutrality, and $T_\rme=T_\rmi$ for simplicity. The electrostatic potential $\delta\Phi$ is solved from the gyrokinetic Poisson (quasineutrality) equation \cite{Lee87}:
\begin{equation}
\label{eq:GTC_poisson}
\frac{en_\rmi}{T_\rmi}(\delta\Phi-\delta\tilde{\Phi})=\delta\bar{n}_\rmi-\delta n_\rme.
\end{equation}
Here, $\delta\tilde{\Phi}$ is the second gyroaverged potential:
\begin{equation}
\delta\tilde{\Phi}=\sum_{\bd{k}}\delta\Phi_{\bd{k}}\Gamma_0\left(b\right)\rme^{\rmi\bd{k}\cdot\bd{R}/\rho_*},
\end{equation}
with $\Gamma_0(b)=I_0(b)\rme^{-b}$,  $I_0$ the modified Bessel function of the first kind, and $b=(k_\perp a)^2/2$. Also, $\delta\bar{n}_\rmi$ is the gyroaveraged ion gyrocenter density perturbation.

In our simulations we adopt the two-spatial-scale approximation that ion markers have uniform density and temperature, while the equilibrium distribution function has the effects from finite density and temperature gradient \cite{Lee87}. This can be done as follows: since $f_0$ is a local Maxwellian, its derivatives can be written as
\begin{eqnarray}
\label{eq:GTC_gradf0}
    \nabla f_0=f_0\left[\frac{\pd \ln n_\rmi}{\pd\psi}+\frac{\pd \ln T_\rmi}{\pd\psi}\left(\frac{\mc{E}}{T_\rmi}-\frac{3}{2}\right)\right]\nabla\psi,\\
    \frac{\pd f_0}{\pd \mc{E}}=-\frac{f_0}{T_\rmi}.
\end{eqnarray}
We assume constant $n_\rmi$ and $T_\rmi$ when evaluating $f_0$ and $\pd_\mc{E}f_0$, but also assume nonzero $\pd_\psi\ln n_\rmi$ and $\pd_\psi\ln T_\rmi$ when evaluating $\nabla f_0$. Consistent with the convention in local simulations, we define the local density and temperature gradient scale lengths $L_n =(aB_a\pd_\psi\ln n_\rmi)^{-1}$ and $L_T= (aB_a\pd_\psi\ln T_\rmi)^{-1}$, and refer to $a/L_n$ and $a/L_T$ as the local gradients. Here,  $a$ is obtained from the quantity Aminor\_p of the VMEC equilibria and $B_a=\psi_a/(\pi a^2)$, where $\psi_a$ is the value of $\psi_{\rm V}$ at the outermost flux surface. For our simulations we set the density gradient to be zero, $a/L_n=0$. The temperature gradient has the following form:
\begin{equation}
    \frac{a}{L_T(r)}=\frac{1}{2}\frac{a}{L_{T0}}\left(\tanh\frac{r/a-0.4}{\Delta r}+\tanh\frac{0.6-r/a}{\Delta r}\right),
\end{equation}
where $r=\sqrt{\psi_{\rm V}/(\pi B_a)}$, $\Delta r=0.02a$, and we choose $a/L_{T0}=2$. Meanwhile, the marker temperature itself is assumed constant, $T_\rmi=T_{\rmi 0}$. Note that nonuniform density and temperature profiles can be used in GTC simulations of actual experiments and that, in the current study, even with the assumption of constant $n_\rmi$ and $T_\rmi$ and their gradients, the geometry remains global, so we do not expect the GTC simulation results to be identical to the local simulation results. 

\begin{figure}
    \centering
    \includegraphics[width=1\linewidth]{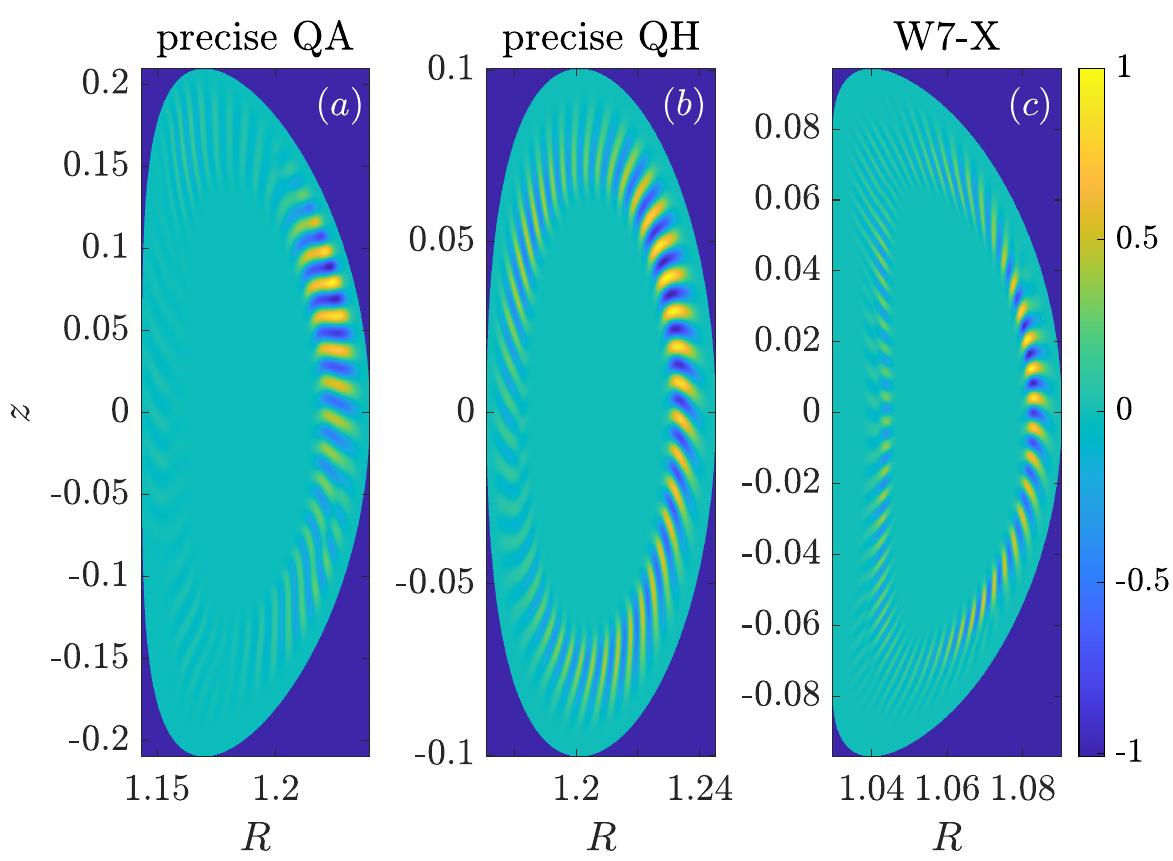}
    \caption{The linear global ITG eigenmode structures ${\rm Re}\,\delta\Phi_{\omega_\rmr}$ calculated from \eref{eq:GTC_delta_Phi} at $\zeta=0$. Lengths $(R,z)$ are normalized by  $R_0$ of each configuration.}
    \label{fig:GTC_linear_Rz}
\end{figure}

Since different configurations have different $B_a$ and $a$, we choose $T_{\rmi0}$ accordingly so that $\rho_*$ are approximately the same. We have $a=0.168~{\rm m}$ and $B_a=0.92~{\rm T}$ for the precise QA, $a=0.125~{\rm m}$ and $B_a=0.78~{\rm T}$ for the precise QH, and $a=0.494~{\rm m}$ and $B_a=2.6~{\rm T}$ for W7-X. Therefore, we use $T_{\rmi 0}=64~{\rm eV}$, $16~{\rm eV}$, and $4000~{\rm eV}$ for the three configurations, respectively, so that $\rho_*\approx 0.01$.  We also choose $n_\rmi=10^{19} ~\rmm^{-3}$, which does not affect the linear eigenmode structures. In our simulations, we use 2000 grids in the poloidal direction (about 40 grids per wave period), 40 radial grids in the domain $0.35<r/a<0.65$, and require $\delta\Phi=0$ at the radial boundaries. We simulate 15 toroidal planes spanning one field period of the devices, $-\pi/N_{\rm fp}<\zeta<\pi/N_{\rm fp}$, where $N_{\rm fp}=2$ for the QA, $N_{\rm fp}=4$ for the QH, and $N_{\rm fp}=5$ for W7-X. Approximately 100 ions are simulated for each grid point in the 3D domain, and the simulation time step size is $\rmd t=0.01 R_0/v_{\rm ti}$ with $R_0$ the toroidally averaged major radius of the magnetic axis. These numerical resolutions have been chosen after careful numerical convergence studies \cite{Wang20,Chen25a}.

\begin{figure}
    \centering
    \includegraphics[width=1\linewidth]{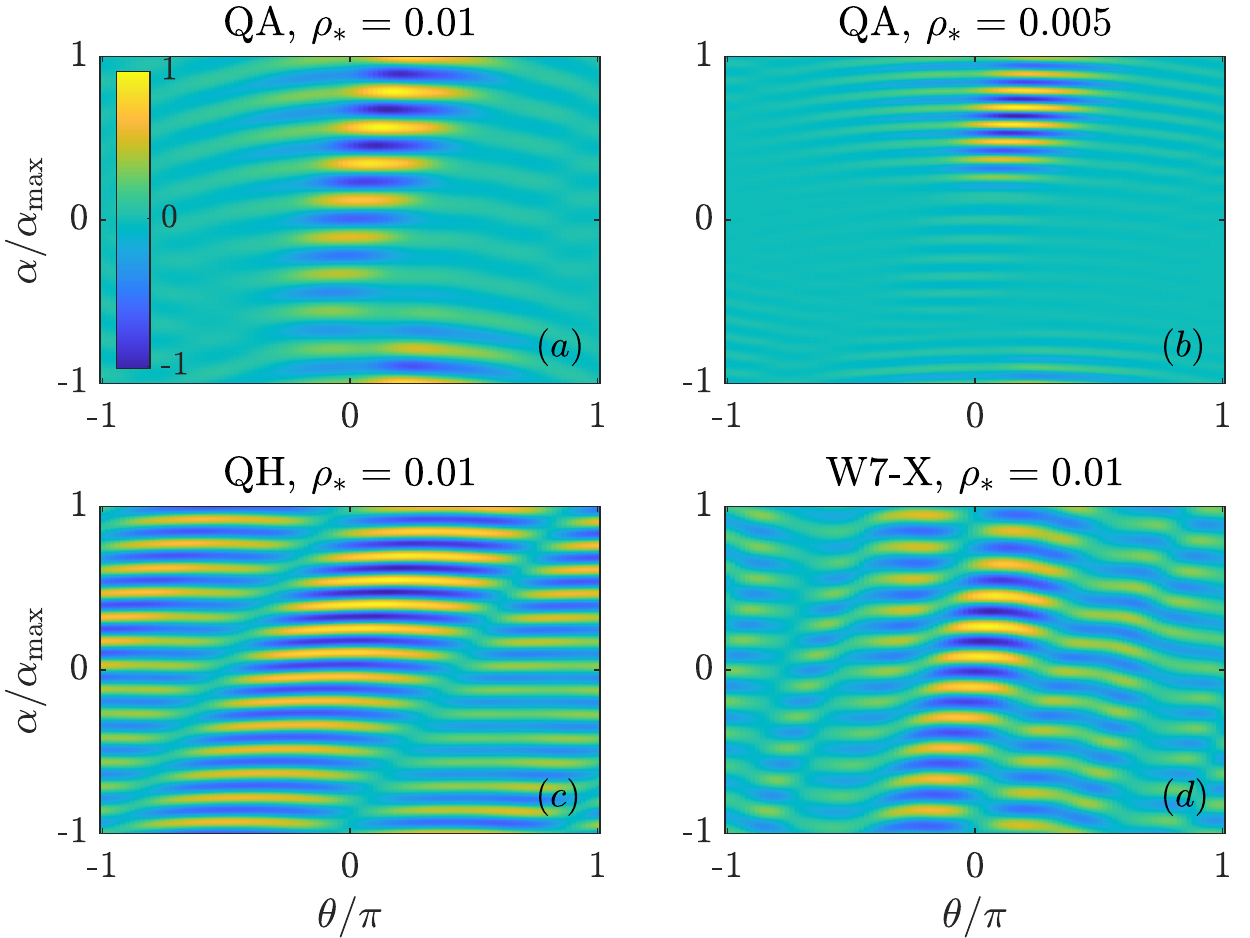}
    \includegraphics[width=1\linewidth]{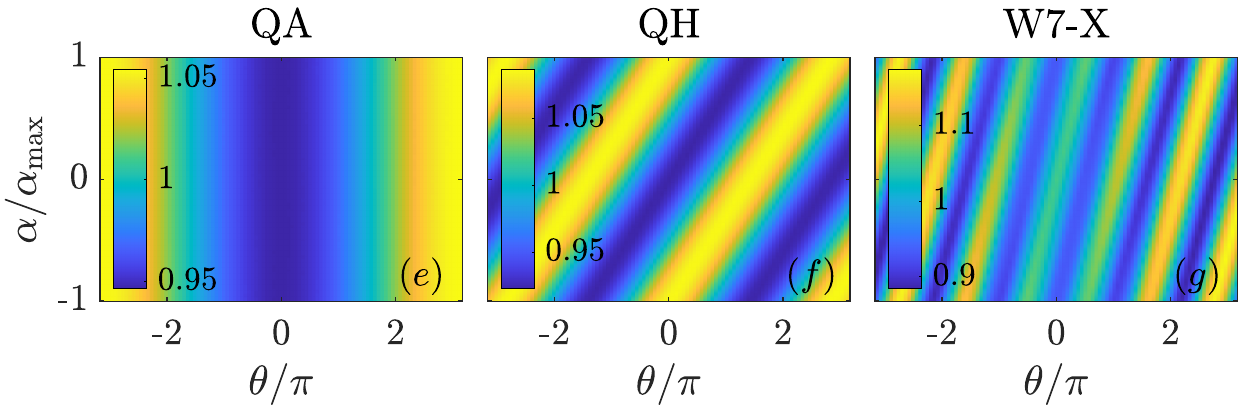}
    \caption{(a-d) The linear ITG eigenmode structures ${\rm Re}\,\delta\Phi_{\omega_\rmr}$ calculated from \eref{eq:GTC_delta_Phi} in field-line following coordinates $(\alpha,\theta)$ at $r/a=0.5$, with $\alpha_{\max}=\iota\pi/N_{\rm fp}$. (e-g) The normalized magnetic-field strength $B/B_a$ versus $(\alpha,\theta)$ at $r/a=0.5$. Note that the range in the colorbar is different from (a-d).}
    \label{fig:GTC_linear_alphatheta}
\end{figure}

We initialize the simulations with random noise in the ion weights, and observe the most unstable ITG eigenmodes emerge. We assume global eigenmode of the from $\delta\Phi\propto\rme^{-\rmi\omega t}$ and calculate the linear growth rate $\gamma={\rm Im}\,\omega$ from the mode amplitude:
\begin{equation}
    |\delta\Phi|^2\propto \rme^{2\gamma t}.
\end{equation}
Then, we obtain the Fourier spectrum with respect to the real frequency $\omega_\rmr={\rm Re}\,\omega$:
\begin{equation}
\label{eq:GTC_delta_Phi}
    \delta\Phi_{\omega_\rmr}(\bd{R})=\int\delta\Phi(\bd{R},t)\rme^{\rmi\omega_\rmr t-\gamma t}\rmd t.
\end{equation}
The eigenmode is identified where $|\delta\Phi_{\omega_\rmr}|$ is maximized. We obtain the most unstable  eigenmode frequencies $\omega a/v_{\rmt\rmi}=0.266+0.108\rmi$ for the QA, $0.183+0.078\rmi$ for the QH, and $0.287+0.069\rmi$ for W7-X. The corresponding eigenmode structures at $\zeta=0$ are plotted in \fref{fig:GTC_linear_Rz}.  The modes rotate counter-clockwise in the direction of ion diamagnetic drift, while their amplitudes increase exponentially in time. Due to stellarator symmetry, the configurations are up-down symmetric at $\zeta=0$. However, the mode structures are up-down asymmetric, and are localized at the upper part of the plane, which is the downstream of the ion diamagnetic drift. 

The up-down asymmetry of the mode structures is a result of the localization in $\alpha$. \Fref{fig:GTC_linear_alphatheta} shows the mode structures as well as $B$ in the field-line following coordinates $(\alpha,\theta)$ on the $r/a=0.5$ flux surface, where we use the poloidal angle $\theta$ instead of the distance along field lines $l$ as the parallel coordinate. The rotational transform at this radial location is $\iota=0.42$ for the QA, $\iota=-1.24$ for the QH, and $\iota=0.88$ for W7-X. Two features arise due to the nonaxisymmetric stellarator geometries. First, for mode structures along field lines, they are not centered at the bad-curvature regions. For example, the bad-curvature region coincides with the minimum of $B$ at $\theta=0$ for QA, but the mode structure does not peak at $\theta=0$ except at $\alpha=0$. This feature is also verified from local simulations in \sref{sec:local}. For W7-X, however, the minimum of $B$ does not necessarily coincide with the bad-curvature region.) Second, for mode structures across field lines, they do not obey stellarator symmetry, which states that geometric quantities are the same at  $(\alpha,\theta)$ and $(-\alpha,-\theta)$. Instead, the modes are localized at $\alpha>0$, which results in the up-down asymmetry seen from \fref{fig:GTC_linear_Rz}. We also found that the localization becomes more pronounced at  smaller $\rho_*$, as can be seen from \fref{fig:GTC_linear_alphatheta}(b), where we did another simulation of the ITG mode in QA at $T_{\rmi0}=16~{\rm eV}$ so that $\rho_*\approx0.005$. (The corresponding normalized eigenmode frequency for this case is $\omega a/v_{\rmt\rmi}=0.246+0.106\rmi$, which is very close to the case with $\rho_*=0.01$.) Similar behaviors at smaller $\rho_*$ are found for the QH and QI, which are not shown here. This behavior will be explained in \sref{sec:theory} below, where we conclude that the localization increases exponentially with $\rho_*^{-1}$.
\section{Theoretical explanation from a simple model}
\label{sec:theory}
Zocco \etal proposed a simple model to study the  mode structures across field lines \cite{Zocco16,Zocco20}. The model considers the limit of large temperature gradient, $a/L_T\gg 1$, so that the ITG mode is concentrated at the bad-curvature region along field lines and the $\theta$-dependence of the mode structure is not considered. Then, the model becomes a 1D equation in $\alpha$ which describes the global eigenmode structures as well as their frequencies. Consider eigenmode structures of the form $\delta{\Phi}=\Phi(\alpha)\rme^{-\rmi\omega t}$, where $\Phi(\alpha)=\hat{\Phi}(\alpha)\rme^{\rmi S(\alpha)/\rho_*}$ consists of a quickly varying phase factor $\rme^{\rmi S/\rho_*}$ and a slowly varying envelope $\hat{\Phi}$. For easier comparisons with numerical results, time and frequency are normalized quantities, $t=t_{\rm phys}v_{\rmt\rmi}/a$ and $\omega=\omega_{\rm phys}a/v_{\rmt\rmi}$. To the lowest order in $\rho_*$, we have $\pd_\alpha\delta{\Phi}\approx \rmi\rho_*^{-1}k_\alpha\delta{\Phi}$ with $k_\alpha=\pd_\alpha S$. (The model also assumes the linear growth rate peaks at  $k_\psi=0$, but as will be shown in \sref{sec:local} below this is not always true.) Then, the ITG eigenmode structures can be solved along each field line, which yields the local dispersion relation. In the fluid ($\omega\gg \iota a/R_0$) and long-wavelength ($ak_\alpha|\nabla\alpha|\ll 1$) limits, the local dispersion relation is \cite{Romanelli89,Biglari89,Plunk14,Rodriguez25}:
\begin{equation}
\label{eq:theory_dispersion}
\omega^2-b\omega_T\omega+\omega_\rmd\omega_T=0.
\end{equation}
Here, $\omega_T=k_\alpha \omega_{T0}/\rho_*$ with $ \omega_{T0}=\rho_\rmi/(2L_T)$ the ion diamagnetic-drift frequency, $\omega_\rmd=k_\alpha\omega_{\rmd 0}/\rho_*$ with $\omega_{\rmd 0}= \rho_\rmi a\pd_\psi B$  the ion drift frequency, and $b=k_\alpha^2b_0$ with $b_0=|a\nabla\alpha|^2/2$ representing the finite Larmor radius (FLR) stabilizing effects from gyroaveraging. The ITG mode is unstable at the bad-curvature region where $\omega_\rmd\omega_T>(b\omega_T)^2/4$.

In the following, we use a rescaled field-line label $y=N_{\rm fp}\alpha/\iota$ so that $y\in[-\pi,\pi]$. The mode structure is then $\Phi(y)=\hat{\Phi}(y)\rme^{\rmi S(y)/\rho_*}$. To describe $\Phi(y)$, Zocco \etal made the substitution $k_y\to-\rmi\rho_*\pd_y$, so that \eref{eq:theory_dispersion} can be replaced by
\begin{equation}
\label{eq:theory_global}
    \left(\omega^2-\rmi\rho_*^3g\omega\pd_y^3-\rho_*^2f\pd_y^2\right)\Phi(y)=0,
\end{equation}
with $g=(\iota/N_{\rm fp})^{3}b_0\omega_{T0}$ and $f=(\iota/N_{\rm fp})^{2}\omega_{\rmd 0}\omega_{T0}$. The rationale is that to the lowest order in $\rho_*$, we have $\pd_y\to \rmi\rho_*^{-1}k_y$, so that \eref{eq:theory_global} reduces to  \eref{eq:theory_dispersion}. However, there are alternative forms of \eref{eq:theory_global} that also reduce to \eref{eq:theory_dispersion} at the lowest order. We present a systematic derivation of an integro-differential equation for $\Phi$ in \ref{sec:app}, which is slightly different from \eref{eq:theory_global}. Nevertheless, the difference is not important because as will be shown below, only the lowest order dispersion relation is required to explain the mode localization. Also, \eref{eq:theory_dispersion} is a very crude approximation and generally does not agree with numerical results from  gyrokinetic simulations, but \eref{eq:theory_global} still correctly captures the mode localization in $y$ (\fref{fig:ITG_toy_phi} below), suggesting that such a feature must be universal.

Equation~\eref{eq:theory_global} can be solved numerically, where $\omega$ and $\Phi$ are found as the eigenvalues and eigenvectors of a square matrix $\Omega$ that satisfies 
\begin{equation}
\label{eq:theory_matrix}
    \Omega^2-G\Omega+F=0,
\end{equation}
with $G$ and $F$ the matrix representations of $\rmi\rho_*^3g\pd_y^3$ and $-f\pd_y^2$. Since matrices $\Omega$ and $G$ do not commute, the eigenvalues of $\Omega_\pm=(G\pm\sqrt{G^2-4F})/2$ are not the eigenvalues of $\Omega$. Nevertheless, for each eigenvalue $\omega$ of $\Omega$, we can use the corresponding eigenvalue of $\Omega_\pm$ as the initial guess, and then iteratively solve for $\omega$ as the eigenvalue of $G-F/\omega$. An example of the solution is shown in \fref{fig:ITG_toy_phi}(a) with $\rho_*=0.05$, $f=1$, $g=1-0.2\cos y$, and the most unstable global eigenmode corresponds to $\omega=0.66+0.89\rmi$. Here, we are mimicking a QS stellarator where $f$ does not depends on $y$ but $g$ does. In particular, $g$ is smaller at $y=0$ than $y=\pi$, and the mode is localized around $y=\pi/2$. Zocco \etal obtained a similar localized solution with a different set of parameters more relevant to W7-X, again suggesting that the localization is a universal feature; however, a comprehensive understanding of this feature is still missing. 

\begin{figure}
\centering
\includegraphics[width=1\linewidth]{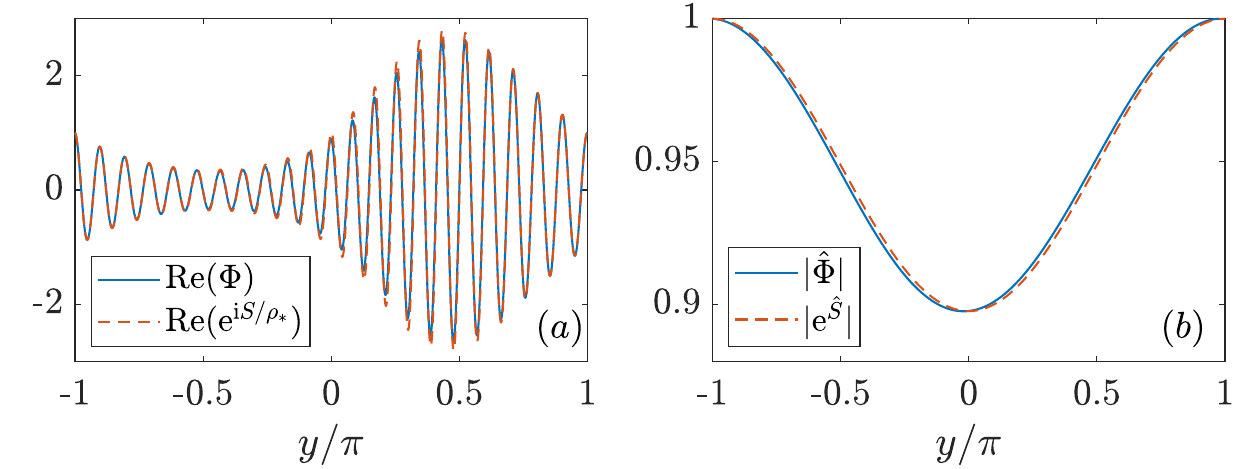}
    \caption{Comparison of numerical and analytical solutions of \eref{eq:theory_global}. Here, $\Phi(\alpha)$ is the numerical solution, $S(\alpha)$ and $\hat{S}(\alpha)$ are analytical results
    \eref{eq:theory_S} and \eref{eq:theory_Shat}, and $\hat{\Phi}(\alpha)=\Phi\rme^{-\rmi S/\rho_*}$. }
    \label{fig:ITG_toy_phi}
\end{figure}

In the following, we study the mode structure from a WKB analysis similar to Dewar and Glasser \cite{Dewar83}. Since $\Phi(y)=\hat{\Phi}(y)\rme^{\rmi S(y)/\rho_*}$, we replace $\pd_y$ with $\rmi \pd_yS/\rho_*+\pd_y$, where $\pd_y$ no longer acts on the phase factor. To the lowest order in $\rho_*$, \eref{eq:theory_global} becomes
\begin{equation}
\label{eq:theory_dispersion2}
    \omega^2- gk_y^3\omega+fk_y^2=0,\quad k_y(y)= \pd_y S,
\end{equation}
which is nothing but the local dispersion relation \eref{eq:theory_dispersion}. To the next order in $\rho_*$, we obtain an equation for the envelope $\hat{\Phi}$:
\begin{equation}
\label{eq:theory_envelope}
    \frac{\pd_y\hat{\Phi}}{\hat{\Phi}}=-\frac{\pd_y k_y}{k_y}\frac{3gk_y\omega-f}{3gk_y\omega-2f}.
\end{equation}
Therefore,  the global solution is  $\Phi=\rme^{\hat{S}}\rme^{\rmi S/\rho_*}$ with
\begin{eqnarray}
    \label{eq:theory_S}
    S(y)=\int_0^y k_y(s)\rmd s,\\
    \label{eq:theory_Shat}
    \hat{S}(y)=-\int_0^y \frac{\pd_sk_y(s)}{k_y(s)}\frac{3g(s)k_y(s)\omega-f(s)}{3g(s)k_y(s)\omega-2f(s)}\rmd s.
\end{eqnarray}
As shown in \fref{fig:ITG_toy_phi}, the numerical solution of $\Phi$ is well approximated by the phase factor $\rme^{\rmi S/\rho_*}$. The numerically calculated envelope $\hat{\Phi}$ also agrees well with the next-order result $\rme^{\hat{S}}$. However, the variation in $|\hat{\Phi}|$ is insignificant compared to the exponentially varying phase factor. Therefore, we focus on the  phase factor in the following.

\begin{figure}
    \centering
    \includegraphics[width=1\linewidth]{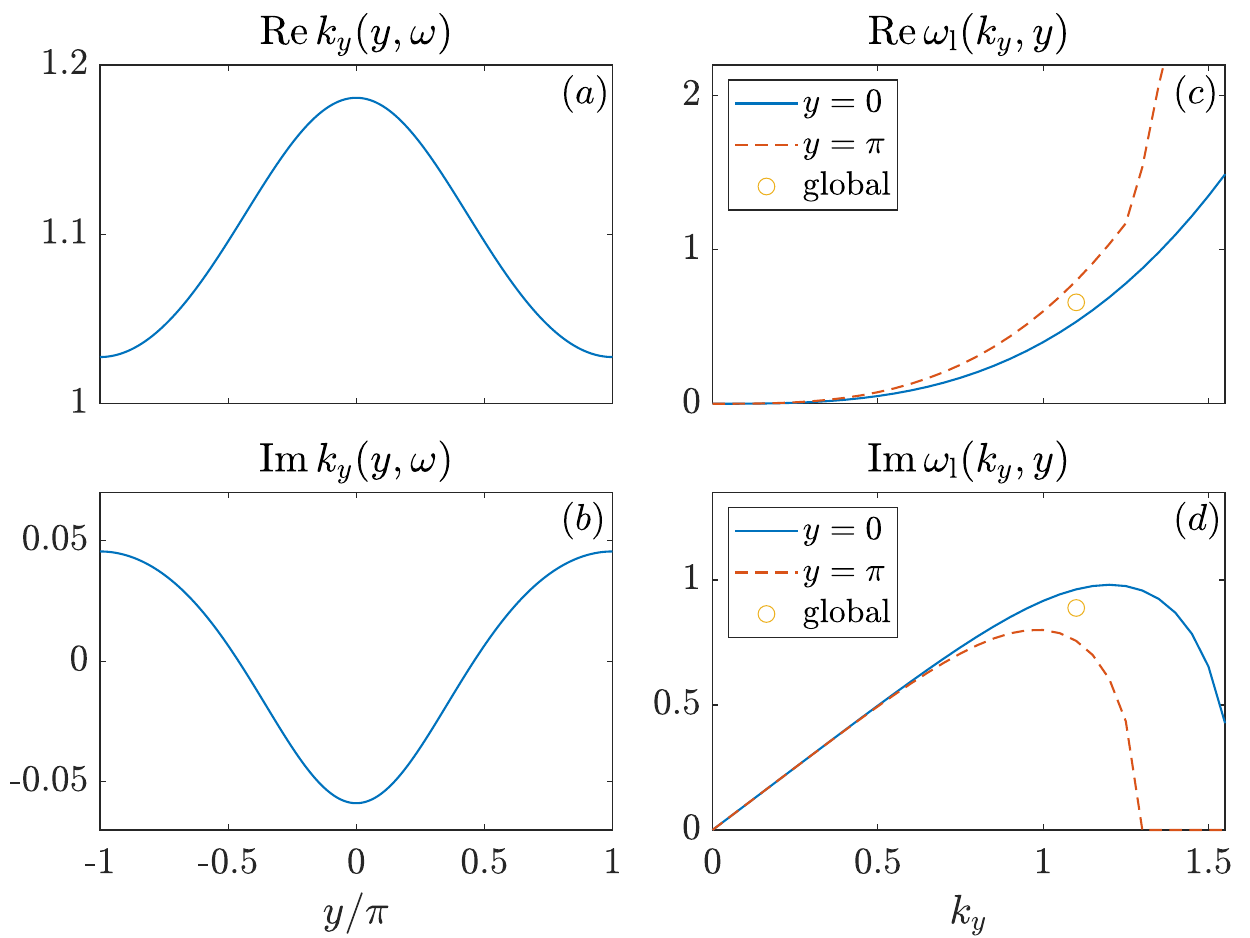}
    \caption{(a) and (b): the real and imaginary part of the global wavenumber versus $y$ at $\omega=0.66+0.89\rmi$. (c) and (d): the local frequency and growth rate versus real $k_y$ at $y=0$ and $y=\pi$. The  global eigenvalue $\omega$ is indicated by the yellow circle.}
    \label{fig:ITG_toy_ky}
\end{figure}

To understand the localization of $\rme^{\rmi S/\rho_*}$ at $y>0$,  we solve for $S$ from \eref{eq:theory_dispersion2} and \eref{eq:theory_S}. Since $\omega$ is constant while $f$ and $g$ vary in $y$ (though $f$ is constant for the example shown here), $k_y$ must vary with $y$ too. Namely, we solve for $k_y(y,\omega)$ at given $\omega$. The solution is shown in \fref{fig:ITG_toy_ky}(a) and (b). The imaginary part of $k_y$ is nonzero, ${\rm Im}\,k_y\approx -0.05\cos y$. Consequently, the phase factor has a nonuniform amplitude,
\begin{equation}
    |\rme^{\rmi S/\rho_*}|\approx \rme^{(0.05\sin y)/\rho_*},
\end{equation}
which is localized around $y=\pi/2$. Therefore, the localization is due to the negative ${\rm Im}\,k_y$ at $y=0$ and increases exponentially with $\rho_*^{-1}$.

To better understand why $k_y$ is complex, and why ${\rm Im}\,k_y<0$ at $y=0$, we solve for the local dispersion relation $\omega_{\rm l}(k_y,y)$ at different $y$ from \eref{eq:theory_dispersion2} and the results are shown in \fref{fig:ITG_toy_ky}(c) and (d). At given $k_y$, the local growth rate ${\rm Im}\,\omega_{\rm l}$ is largest at $y=0$ and smallest at $y=\pi$. This is a common situation for ITG modes in stellarators, where the $y=0$ flux tube gives the largest grow rate. Also shown in \fref{fig:ITG_toy_ky}(c) and (d) are the global eigenmode frequency $\omega$, which is somewhere between the local results. When we solve for $k_y=k_y(y,\omega)$ from $\omega_{\rm l}(k_y,y)=\omega$, the solution cannot be found on the real-$k_y$ axis, and hence $k_y$ must be complex and can be approximately obtained from the first-order Taylor expansion:
\begin{equation}
\label{eq:theory_taylor}
    k_y\approx k_{y 0}+\left(\frac{\pd\omega_{\rm l}}{\pd k_y}\right)^{-1}\Delta\omega,\quad \Delta\omega=\omega-\omega_{\rm l},
\end{equation}
with $\pd \omega_{\rm l}/\pd k_y$ evaluated at $k_y=k_{y 0}$. We choose $k_{y 0}=1.1$ as the averaged ${\rm Re}\, k_y$ from \fref{fig:ITG_toy_ky}(a), but its exact value is not important as long as it is close to $k_y$ so that \eref{eq:theory_taylor} is valid. Since we are looking at the most unstable eigenmode where $\pd({\rm Im}\,\omega)/\pd k_y\approx 0$,  the sign of ${\rm Im}\,k_y$ is determined by the sign of $[\pd({\rm Re}\,\omega)/\pd k_y]^{-1}{\rm Im}\,\Delta\omega$. Therefore, we identify two reasons that lead to the negative ${\rm Im}\, k_y$ at $y=0$: (i) the global eigenmode growth rate is smaller than the local growth rate at $y=0$, ${\rm Im}\,\Delta\omega<0$; and (ii) the ion diamagnetic drift is defined to be in the positive-$y$ direction, so that $\pd({\rm Re}\,\omega)/\pd k_y>0$. In other words, if the $y=0$ flux tube gives the largest local ITG growth rate, then the localization of the ITG modes will occur downstream of the ion diamagnetic drift.

The fact that the global growth rate is smaller than the local growth rate at $y=0$ has already been mathematically proved in \cite{Zocco16,Zocco20}, but our approach here provides a more transparent explanation. Since the solution must be periodic in $y$, we require $\int_{-\pi}^{+\pi}\rmd y\,k_y/\rho_*=2\pi n$ with an integer $n$. For the real part, this requirement can always be satisfied because $\rho_*$ can be arbitrarily chosen. For the imaginary part, this requires $\int_{-\pi}^{+\pi}\rmd y\,{\rm Im}\,\Delta\omega\approx 0$, so that the global growth rate must be somewhere between the local values at $y=0$  and $y=\pi$. A physical interpretation is also given in \cite{Helander15} that the wave experiences different growth rates (assuming real $k_y$) as it propagates across field lines, so the overall growth rate is flux-surface averaged. Our interpretation is different in that $\omega$ (hence the growth rate) remains the same across field lines, \ie an eigenvalue of a global mode, but $k_y$ is allowed to be complex.

Finally, we note that the above conclusions are different from the results from Dewar and Glasser \cite{Dewar83,Dewar97} who studied the global ballooning mode structures using ideal magneto-hydrodynamic (MHD) equations. Since the ideal MHD is self-adjoint, $k_y$ must be purely real or purely imaginary from the local dispersion relation, so that the phase factor cannot lead to localization in $y$, and the next-order envelope equation must be considered to capture the spatial distribution of the mode structure.
\section{Construction of a surface-global solution from local gyrokinetic simulation results}
\label{sec:local}
Based on the results from \sref{sec:theory}, we construct a surface-global solution from local gyrokinetic simulations, and compare the results with the GTC solution. In particular, we focus on the precise QA configuration here. First, we obtain the local dispersion relation $\omega=\omega_{\rm l}(\bd{k},\alpha)$ with real $\bd{k}=k_r\nabla r+k_\alpha\nabla\alpha$. Then, we calculate the complex $\bd{k}$ from first-order Taylor expansion, and construct a surface-global solution using the complex $\bd{k}$. Finally, we compare the constructed solutions with the GTC solutions. Following the convention in local simulations, we use  normalized eigenmode frequencies $\omega a/v_{\rm ti}$ with $v_{\rmt\rmi}=\sqrt{2T_\rmi/m_\rmi}$, and  normalized radial and binormal wavenumbers $k_x=k_r a$ and $k_y=k_\alpha a /r$. Note that the definition of $k_y$ here differs from \sref{sec:theory} by a factor of  $(\iota/N_{\rm fp})(a/r)$. 

\begin{figure}
    \centering
    \includegraphics[width=1\linewidth]{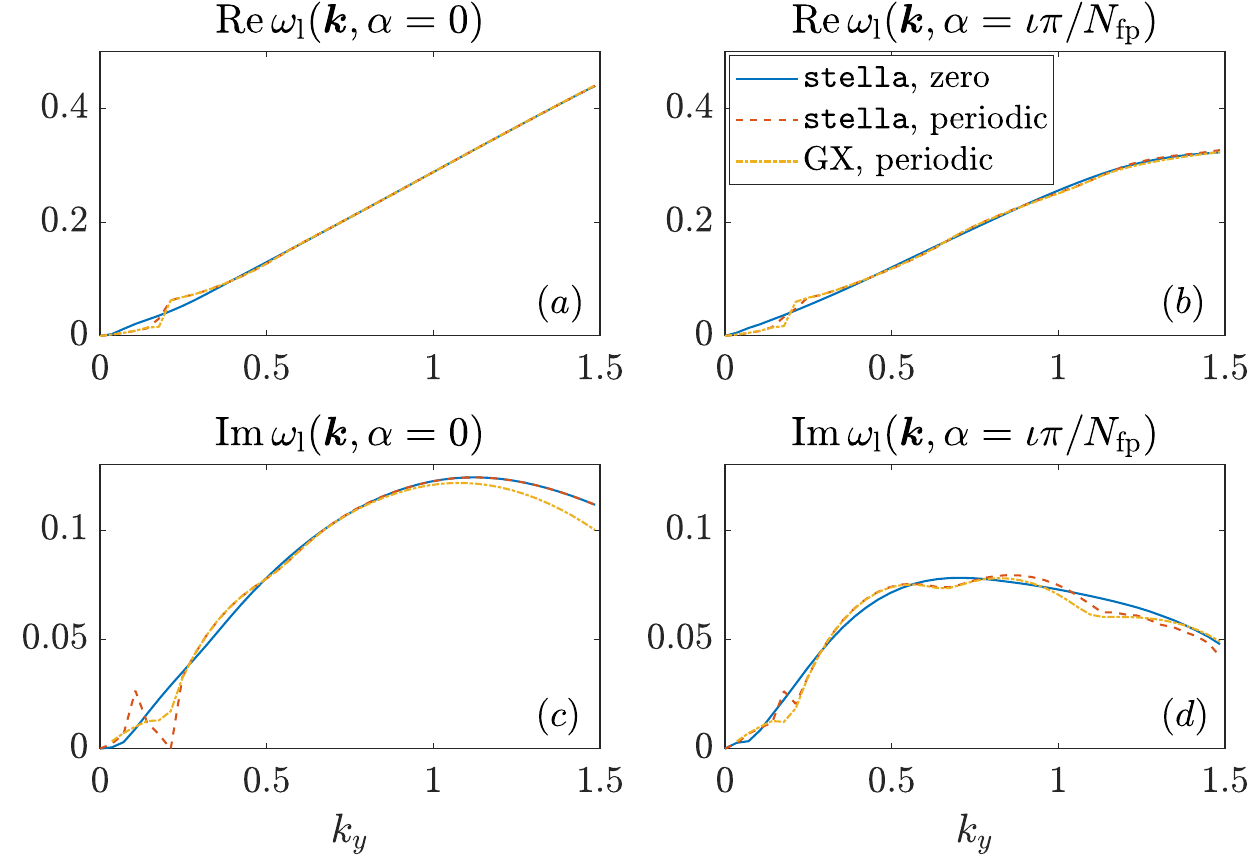}
    \caption{The local linear eigenmode real frequencies (a,b) and growth rates (c,d) versus $k_y$ at $k_x=0$  at two different field lines: $\alpha=0$ and $\alpha=\iota\pi/N_{\rm fp}$. Shown are results from \texttt{stella} using the zero-incoming boundary condition, and from both \texttt{stella} and GX using the periodic boundary condition.}
    \label{fig:local_1d}
\end{figure}

We use local gyrokinetic codes \texttt{stella} \cite{Barnes19} and GX \cite{Mandell24} to obtain $\omega_\rml(\bd{k},\alpha)$ at the $r/a=0.5$ flux surface. Here, we compare the results from the two codes so that we can evaluate the effects from parallel boundary conditions (\fref{fig:local_1d}). The two codes use different sign conventions: $\bd{B}=\nabla\psi\times\nabla\alpha$ in GX but $\bd{B}=\nabla\alpha\times\nabla\psi$ in \texttt{stella}. Therefore, assuming the same $\psi$, the field line $\alpha$ in GX corresponds to the field line $-\alpha$ in \texttt{stella}, and the real frequencies of ITG modes ${\rm Re}\,\omega_\rml$ are positive in GX but negative in \texttt{stella}. To be consistent with the sign convention in \sref{sec:GTC} and \sref{sec:theory}, we will flip the signs of $\alpha$, $k_x$, and ${\rm Re}\,\omega_\rml$ for the \texttt{stella} results presented in this section. For each field line $\alpha$, we choose the flux tube to span one poloidal turn. The flux tubes in GX are centered at $\theta=0$ so that $\theta\in(-\pi,\pi)$, while the flux tubes in \texttt{stella} are centered at $\zeta=0$ so that $\theta\in(-\pi+\alpha,\pi+\alpha)$. For the local simulations, $\delta\!f$ is decomposed into Fourier components:
\begin{equation}
\delta\!f=\sum_{\bd{k}}\delta\!f_{\bd{k}}(\bd{k},\theta,\mu,\mc{E},\sigma,t;\alpha)\rme^{\rmi\bd{k}\cdot\bd{R}/\rho_*},
\end{equation}
where $\sigma=v_\parallel/|v_\parallel|$ and $\alpha$ is treated as a parameter. For electrostatic collisionless simulations, local codes solve the same linear gyrokinetic equation as \eref{eq:GTC_linear}, but with the eikonal assumption $\nabla\to\rmi\bd{k}/\rho_*+\nabla_\parallel$, so that
\begin{equation}
    \nabla\delta\!f=\sum_{\bd{k}}(\rmi\bd{k}/\rho_*+k_\parallel\pd_\theta)\delta\!f_{\bd{k}}\rme^{\rmi\bd{k}\cdot\bd{R}/\rho_*},
\end{equation}
where $k_\parallel=(\pd l/\pd\theta)^{-1}$.  Therefore,  local codes solve $\delta\!f_{\bd{k}}$ from an ordinary differential equation in $\theta$ for each $\bd{k}$. The potential $\delta\Phi$ is also decomposed into the Fourier basis, and is solved from the same quasineutrality condition \eref{eq:GTC_poisson}. The local eigenmode frequency $\omega_\rml(\bd{k},\alpha)$ is then found by fitting $\delta\!f_{\bd{k}}$ to  $\rme^{-\rmi\omega_\rml t}$. Since the eikonal assumption is valid to the lowest order in $\rho_*$, global simulations can be interpreted as solving the ordinary differential equations along field lines for each $\bd{k}$ at this order. However, due to nonaxisymmetry, $\bd{k}$ can be complex in this interpretation, although global simulations do not actually implement the decomposition in $\bd{k}$. In contrast, local simulations assume periodic boundary conditions perpendicular to field lines, so that $\bd{k}$ must always be real.

\begin{figure*}
    \centering
    \includegraphics[width=1\linewidth]{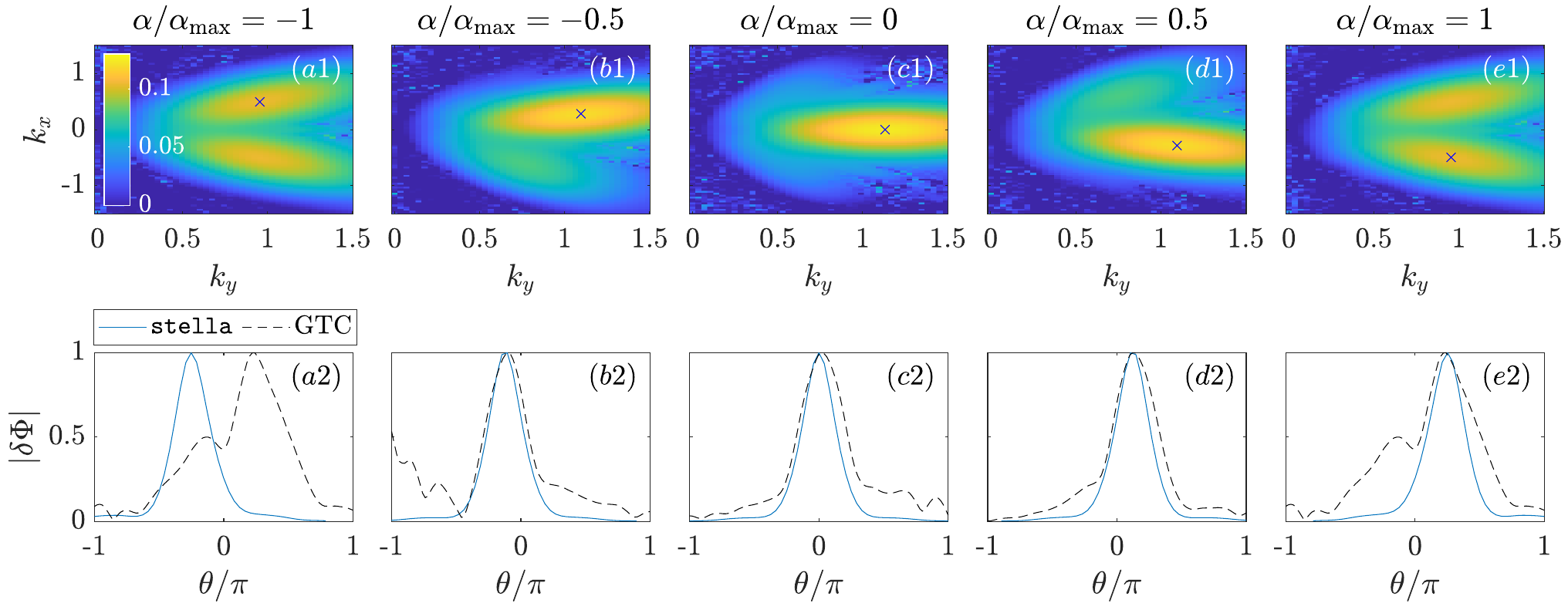}
    \caption{Comparison between results from local \texttt{stella} simulations and global GTC simulations. First row: the local growth rates ${\rm Im}\,\omega_{\rml}$ versus $(k_x,k_y)$ at different $\alpha$, with $\alpha_{\max}=\iota\pi/N_{\rm fp}$. The blue crosses indicate the most unstable local solution at $\bd{k}=\bd{k}_0$ \eref{eq:local_k0}. Second row: the eigenmode amplitudes $|\delta\Phi|$ versus $\theta$, which are normalized to their maxima at each $\alpha$. The \texttt{stella} results are at $\bd{k}=\bd{k}_0$ and the GTC results are from \fref{fig:local_2d}(b). The results at $\alpha/\alpha_{\rm max}=\pm 1$ are identical due to periodicity of the system, where $|\delta\Phi|$ from GTC has two peaks and appears to be a superposition of two local solutions.}
    \label{fig:local_2d}
\end{figure*}

The choice of the parallel boundary condition requires further considerations. The generalized twist-and-shift boundary condition \cite{Martin18}  connects Fourier modes with different $k_x$ together, so they have the same $\omega_\rml$, which is not physical because the two ends of the flux tubes are not physically connected. When the temperature gradient is large, $R_0/L_T\gg 1$, and when $k_y /\iota^2$ is not too small, the ITG mode belongs to the toroidal branch  and is concentrated in the bad-curvature region \cite{Plunk14,Rodriguez25}. Since we have chosen  $a/L_T=2$ in \sref{sec:GTC} and $R_0/a\approx 6$ for the precise QA configuration, the criterion $R_0/L_T\gg 1$ is well satisfied. Also, as will be shown in \fref{fig:local_2d}, $k_y\approx 1$ for the fastest growth ITG eigenmodes, so that they belong to the toroidal branch and decay to almost zero at the good-curvature regions $\theta=\pm\pi$, as can be verified from \fref{fig:GTC_linear_alphatheta}(a) and (b). Then, the zero-incoming boundary condition is more appropriate, which requires zero incoming perturbed distribution at the parallel boundaries \cite{Kotschenreuther95}:
\begin{equation}
    \delta\!f_{\bd{k}}(\theta=\theta_{\min},\sigma=1)=\delta\!f_{\bd{k}}(\theta=\theta_{\max},\sigma=-1)=0.
\end{equation}
To verify that the results are insensitive to boundary conditions where  the mode amplitudes are close to zero, we also test the self-periodic boundary condition:
\begin{equation}
        \delta\!f_{\bd{k}}(\theta=\theta_{\min})=\delta\!f_{\bd{k}}(\theta=\theta_{\max}).
\end{equation}
Simulation results from \texttt{stella} and GX with the two boundary conditions are shown in \fref{fig:local_1d}. For both codes, the resolution is 48 in $\theta$, 8 in $\mu$, and 16 in $v_\parallel$, and the simulation time step is $\rmd t=0.05a/v_{\rmt\rmi}$. The two boundary conditions produce qualitatively the same results, and the results from \texttt{stella} with  zero-incoming boundary condition will be presented in the following. There are also quantitative differences between \texttt{stella} and GX in the growth rates, possibly because the ranges in $\theta$ of the flux tubes are not identical, but the eigenmode structures look the same. One could also try to avoid the issue from parallel boundary conditions by extending the flux tubes to multiple poloidal turns. However, since the modes are concentrated at the bad-curvature region and the simulated stellarator configurations have very low global magnetic shear, the results will be the same as the simulations of multiple flux tubes, each spanning one poloidal turn.

After obtaining the local dispersion relation with real $\bd{k}$,  we look for complex $\bd{k}(\alpha)$ such that $\omega_\rml(\bd{k}(\alpha),\alpha)=\omega$ is constant. Similar to \sref{sec:theory}, this is done from first-order Taylor expansion:
\begin{equation}
\label{eq:local_taylor}
    \bd{k}(\alpha)\approx\bd{k}_0(\alpha)+\left(\frac{\pd\omega_{\rml}}{\pd\bd{k}}\right)^{-1}\Delta\omega,\quad\Delta\omega=\omega-\omega_{\rml}.
\end{equation}
Here, $\bd{k}_0=(k_{x0},k_{y0})$ is real, $\omega$ is the surface-global eigenmode frequency to be determined in the following, and $\omega_\rml$ and $\pd_{\bd{k}}\omega_\rml$ are evaluated at $\bd{k}=\bd{k}_0$. The exact value of $\bd{k}_0$ is not important as long as it is close to $\bd{k}$ so that \eref{eq:local_taylor} is valid. Note that both $\bd{k}$ and $\pd_{\bd{k}}\omega_\rml$ are 2D vectors and  numerically we look for $\Delta\bd{k}=\bd{k}-\bd{k}_0$ that minimizes $|\Delta\bd{k}\cdot\pd_{\bd{k}}\omega_{\rml}-\Delta\omega|$.  In other words, although local simulations only provide the information on the real-$\bd{k}$ plane, we can still obtain the information on the complex-$\bd{k}$ plane from analytic continuation.  

For the choice of $\bd{k}_0$, we search for the most unstable local eigenmodes at each $\alpha$. The local growth rates ${\rm Im}\,\omega_\rml$ versus $(k_x,k_y)$  from \texttt{stella} are plotted in the first row of \fref{fig:local_2d}, which shows that the fastest growing modes correspond to nonzero $k_x$ (in contrast to \sref{sec:theory} where only $k_y$ has been considered). The most unstable modes are marked by the blue crosses in the figure, and are approximately given by
\begin{equation}
\label{eq:local_k0}
k_{x0}=-0.5\alpha/\alpha_{\max},\quad k_{y0}=1.13-0.175(\alpha/\alpha_{\max})^2.
\end{equation}
The mode amplitudes $|\delta\Phi(\theta)|$ at $\bd{k}_0$ are shown in the second row of \fref{fig:local_2d}, where we also plot the GTC results. Due to stellarator symmetry, the local results are symmetric with respect to the coordinate change $(\alpha,\theta,k_x)\to(-\alpha,-\theta,-k_x)$.    As shown in the figure,  the  local eigenmode structures  qualitatively resemble the GTC results. In particular, local results confirm the observation from \fref{fig:GTC_linear_alphatheta}(b) that the mode structures are not centered at $\theta=0$. Note that the GTC solution is periodic in $\alpha$, but the local solution \eref{eq:local_k0} is not, which transitions from $k_{x0}=0.5$ at $\alpha/\alpha_{\max}=-1$ to $k_{x0}=-0.5$ at $\alpha/\alpha_{\max}=1$. In fact, the GTC solution of $|\delta\Phi|$ has two peaks at $\alpha/\alpha_{\max}=\pm 1$, which appears to be a superposition of the two local solutions at $k_{x0}=\pm 0.5$. Therefore, the local solution lives on an extended $\alpha$ space, $\alpha\in(-\infty,\infty)$, and the periodicity of the global solution can be recovered from a superposition of the local solutions \cite{Dewar83}.  

\begin{figure}
    \centering
    \includegraphics[width=1\linewidth]{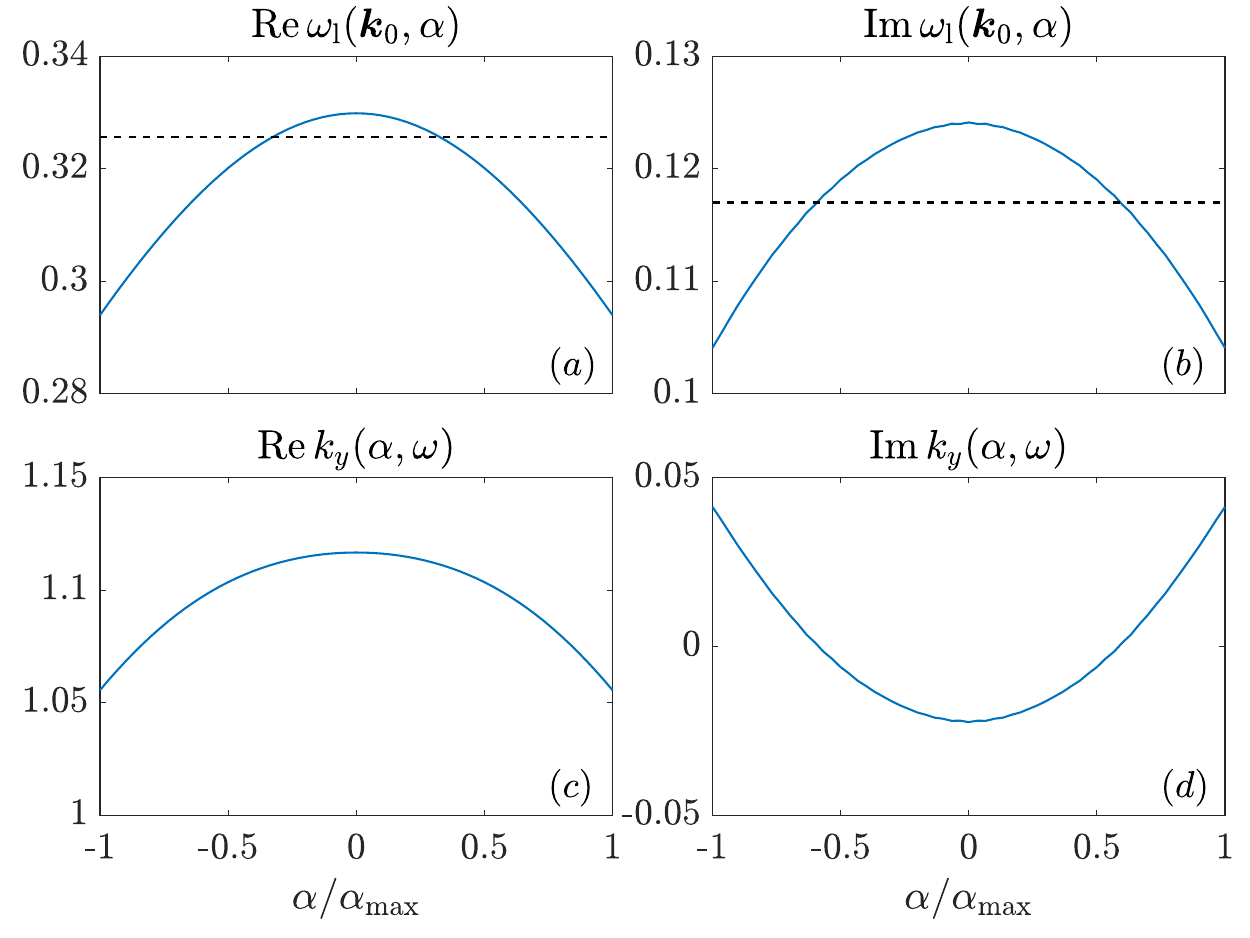}
    \caption{(a) and (b): the real and imaginary parts of $\omega_\rml$ versus $\alpha$ at $\bd{k}=\bd{k}_0$ from \texttt{stella} (blue solid curves) and the calculated surface-global eigenmode frequency $\omega$  (black dashed line). (c) and (d): solution of $k_y$ from first-order Taylor expansion \eref{eq:local_taylor}.}
    \label{fig:local_wavenumber}
\end{figure}

The local eigenmode real frequencies and growth rates  at $\bd{k}=\bd{k}_0$ are plotted in \fref{fig:local_wavenumber}(a) and (b). To determine the surface-global eigenmode frequency $\omega$, we assume that the corresponding solution $\bd{k}$ from \eref{eq:local_taylor} satisfies $\int\rmd\alpha\,{\rm Im}\,k_y=0$. Such constraint comes from the periodicity requirement in \sref{sec:theory}, but here the local solution is not periodic in $\alpha$. In fact, from \fref{fig:local_2d}, the two peaks in the GTC solution of $|\delta\Phi|$ at $\alpha/\alpha_{\max}=\pm 1$ have different amplitudes, so the periodicity constraint is not entirely true. Nevertheless, we use this constraint due to the lack of a better choice to determine $\omega$, and we found that $\omega=0.326+0.117\rmi$. The corresponding solutions for $k_y$ are shown in \fref{fig:local_wavenumber}(c) and (d). In particular, \fref{fig:local_wavenumber}(d) shows that ${\rm Im}\,k_y$ is nonzero and is negative at $\alpha=0$, which leads to the localization  at $\alpha>0$. 

\begin{figure}
    \centering
    \includegraphics[width=1\linewidth]{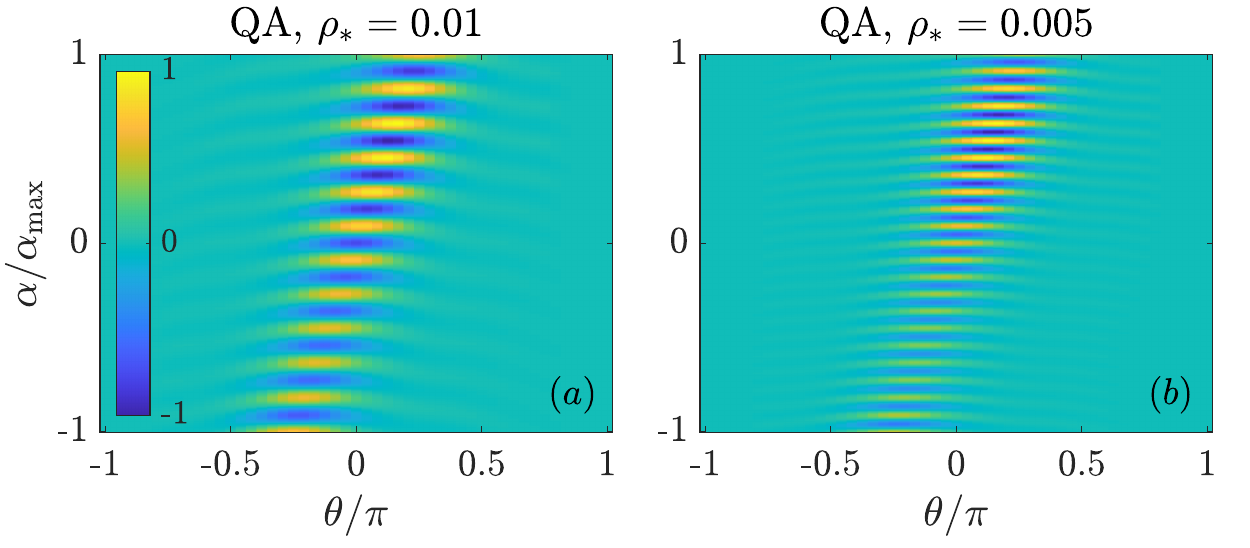}
    \caption{The constructed surface-global solution \eref{eq:local_construct} from local \texttt{stella} solutions with (a) $\rho_*=0.01$ and (b) $\rho_*=0.005$.}
    \label{fig:local_alphatheta}
\end{figure}

Finally, with the complex $\bd{k}$, we construct a surface-global solution  as
\begin{eqnarray}
\label{eq:local_construct}
    \delta\Phi_{\rm g}(\theta,\alpha)=\delta\Phi_\rml(\theta;\alpha,\bd{k}_0)\rme^{\rmi S(\alpha)/\rho_*},
    \\
    S=\int\rmd\alpha\, k_y (\alpha)r/a,
\end{eqnarray}
where $\delta\Phi_\rml$ is the normalized eigenmode structure from \texttt{stella}  at $\bd{k}=\bd{k}_0$ for each $\alpha$. The results are shown in \fref{fig:local_alphatheta} with $\rho_*=0.01$ and $\rho_*=0.005$, which are not periodic in $\alpha$ because we only consider one period in $\alpha$ for the local solution. Nevertheless, the constructed solutions look similar to the  GTC results in \fref{fig:GTC_linear_alphatheta}. The increasing level of localization with decreasing $\rho_*$ is also reproduced as a direct consequence from the phase factor in \eref{eq:local_construct}. Quantitative differences do exist from the GTC results: for the constructed solution here the eigenmode frequency is $\omega=0.326+0.117\rmi$ and the averaged binormal wavenumber is $k_y=1.1$. In comparison, for the GTC solution at $\rho_*=0.01$ shown in \fref{fig:GTC_linear_alphatheta}(a), the eigenmode frequency is $\omega=0.266+0.108\rmi$ and the mode structure is dominated by $k_y=0.95$. While the growth rates are similar, the GTC solution has a lower $k_y$, and hence lower real frequency. From \fref{fig:local_1d}(c) it is seen that the growth rate peaks at a smaller $k_y$ in GX compared to \texttt{stella}. Therefore, if we construct the surface-global solution from GX instead, the resulting $\omega$ will be closer to GTC.

\begin{figure}
    \centering
    \includegraphics[width=1\linewidth]{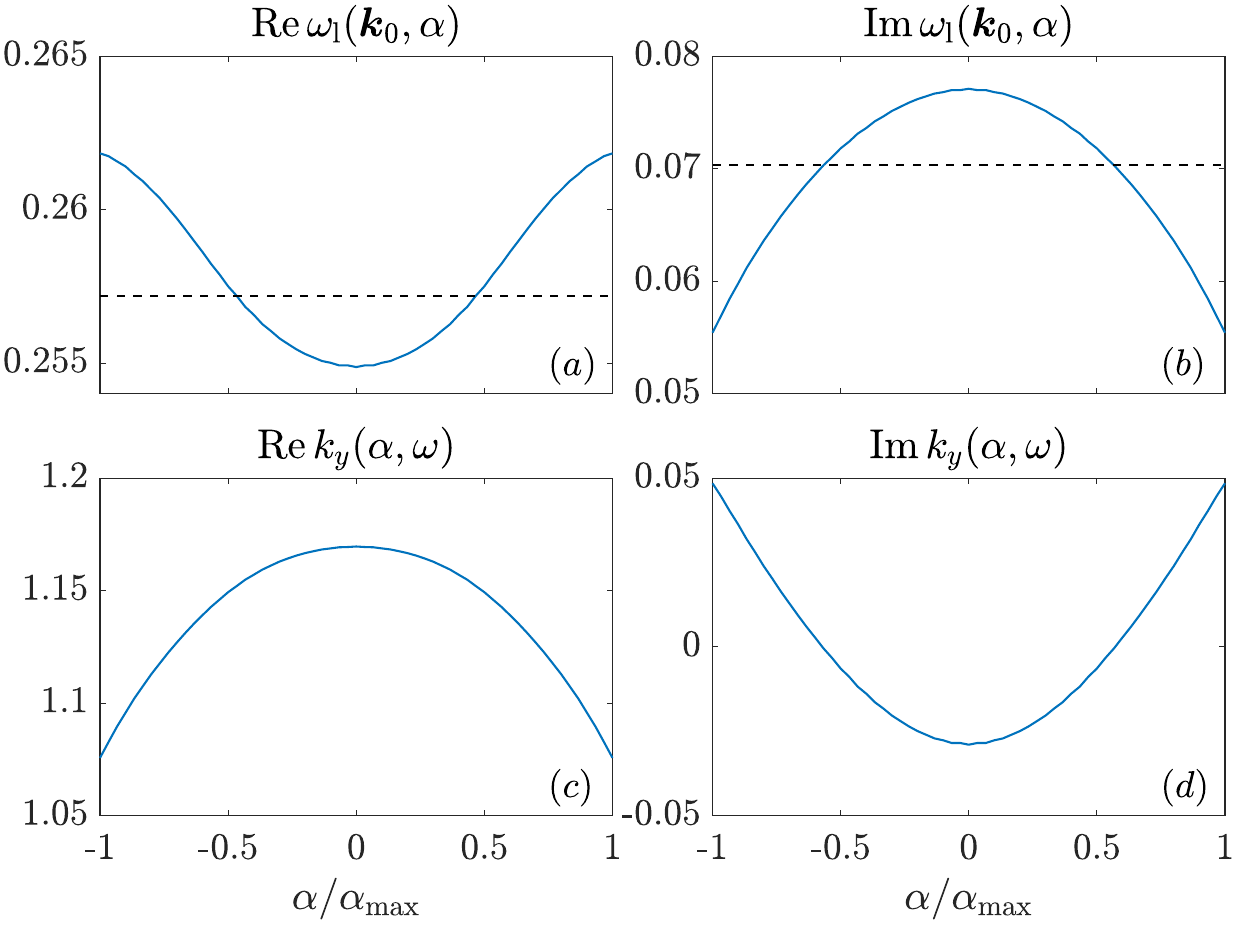}
    \caption{Same as \fref{fig:local_alphatheta} but for the W7-X.}
    \label{fig:local_wavenumber_w7x}
\end{figure}

\begin{figure}
    \centering
    \includegraphics[width=1\linewidth]{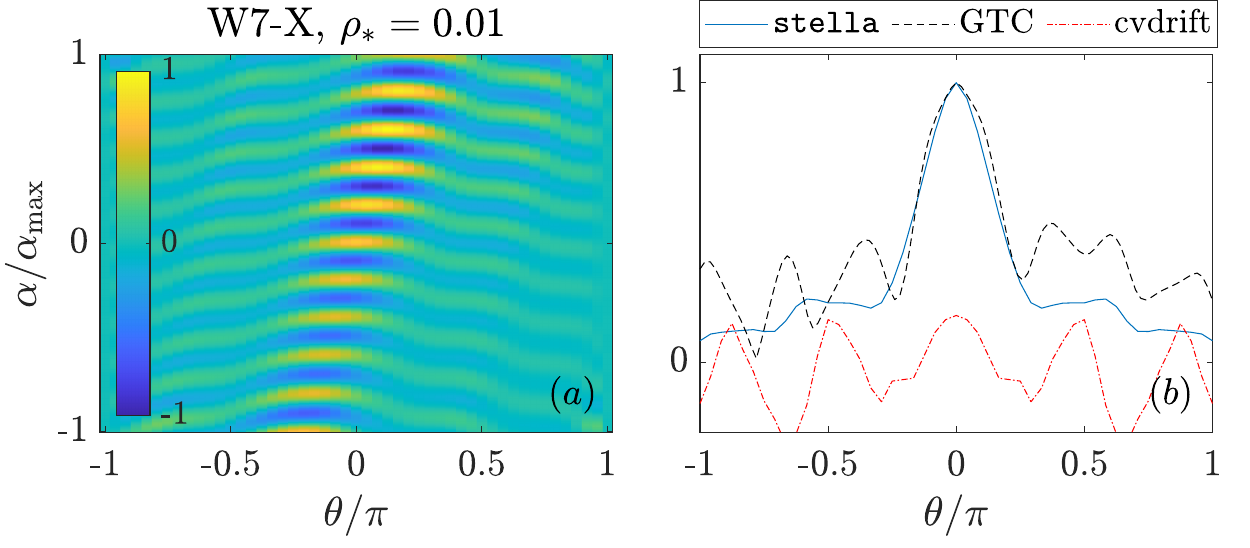}
    \caption{(a) The constructed surface-global solution for the W7-X at $\rho_*=0.01$. (b) The eigenmode amplitudes $|\delta\Phi|$ versus $\theta$ at $\alpha=0$ from \texttt{stella} and GTC simulations. Also shown is the quantity cvdrift=$arB_0^{-1} \bd{B}\times\bd{\kappa}\cdot\nabla\alpha$  along the field line, which is positive in the bad-curvature regions.}
    \label{fig:local_alphatheta_w7x}
\end{figure}

The above analysis has also been carried out for the W7-X configuration using \texttt{stella}. As $\alpha$ varies, the local growth rates change similarly to the precise QA shown in \fref{fig:local_2d}, and the most unstable local eigenmodes $\bd{k}_0$ are approximately given by
\begin{equation}
    k_{x0}=-0.57\alpha/\alpha_{\max},\quad k_{y0}=1.16-0.04(\alpha/\alpha_{\max})^2.
\end{equation}
The local eigenmode real frequencies and growth rates at $\bd{k}=\bd{k}_0$ are plotted in \fref{fig:local_wavenumber_w7x}(a) and (b). The surface-global eigenmode frequency $\omega$ is determined from the same procedure as above and we found that $\omega=0.257+0.070\rmi$, which is  close to the GTC result $0.287+0.069\rmi$. The complex $\bd{k}$ is obtained from first-order Taylor expansion and the real and imaginary parts of $k_y$ are plotted in \fref{fig:local_wavenumber_w7x}(c) and (d). Finally, the constructed surface-global solution is shown in \fref{fig:local_alphatheta_w7x}(a). Note that the constructed solution only includes one period in $\alpha$ and hence cannot describe the entire mode structure on the surface in \fref{fig:GTC_linear_alphatheta}(d), but it does reproduce the dominant structure around $\theta=0$. Also, for the W7-X there are multiple bad-curvature regions within one poloidal turn, and the mode is not completely localized within one bad-curvature region. This is illustrated in \fref{fig:local_alphatheta_w7x}(b), where we plot the quantity $arB_0^{-1} \bd{B}\times\bd{\kappa}\cdot\nabla\alpha$ with $\bd{\kappa}=\hat{\bd{b}}\cdot\nabla\hat{\bd{b}}$  the curvature vector of the magnetic field line. This quantity is commonly named ``cvdrift'' in local codes, and for the $\alpha=0$ field line it has five peaks at $\theta\in(-\pi,\pi)$, corresponding to five bad-curvature regions. (In comparison, cvdrift is proportional to $\cos\theta$ for the precise QA configuration, corresponding to only one bad-curvature region.) Nevertheless, the mode amplitudes still quickly decay as $|\theta|$ increases, so that factors such as the parallel boundary condition and the number of poloidal turns do not play a significant role.

We conclude that for the linear ITG mode structures in the stellarator configurations, many features of the global GTC solution can be reproduced from the constructed surface-global solution, if we first solve the local dispersion relation with real $\bd{k}$, and then do an analytic continuation to the complex $\bd{k}$.

\section{Conclusions and discussions}
\label{sec:conclusion}
We numerically simulate the linear electrostatic ITG eigenmodes in stellarators using the global gyrokinetic particle-in-cell code GTC, and present a theoretical explanation for the observed  mode structures. We find that the linear  eigenmode structures  are localized downstream of the ion diamagnetic drift. Based on a simple model from Zocco \etal \cite{Zocco16,Zocco20} and following the WKB theory of Dewar and Glasser \cite{Dewar83}, we show that the localization can be explained from the nonzero imaginary part of $k_\alpha$. Focusing on the precise QA configuration, we further demonstrate that a localized surface-global eigenmode can be constructed from local gyrokinetic codes \texttt{stella} and GX, only if we first solve the local dispersion relation with real wavenumbers, and then do an analytic continuation  to the complex-wavenumber plane. These results suggest that the complex-wavenumber spectra from surface-global effects are required to understand the linear drift-wave eigenmode structures in stellarators (while existing local simulations always assume real wavenumbers).

While the above conclusions are limited to the linear instabilities, they could be useful in interpreting the nonlinear results.  For example, the nonlinear fluctuation level of ITG turbulence in W7-X deviates from stellarator symmetry \cite{BanonNavarro20,Sanchez23} and its localization is consistent with (although not as pronounced as) the linear results. The linear and nonlinear thresholds for the ITG turbulence in a QA stellarator appears to lie above that of the most unstable flux tube \cite{Xanthopoulos16}, which is also consistent with the conclusion that the global eigenmode growth rate is below the most unstable flux tube. Quantitative studies of these effects will be the subject of future work.

\ack
H.Z. thanks R. Gaur for providing scripts that calculate flux-tube quantities from VMEC equilibrium \cite{Gaur23,Gaur24}. H.Z. was supported by a grant from the Simons Foundation/SFARI (Grant \#560651, AB). This work is supported by the U.S. Department of Energy Award No. DE-SC0024548. This research used resources of the National Energy Research Scientific Computing Center, which is supported by the Office of Science of the U.S. Department of Energy under Contract No. DE-AC02-05CH11231.
\section*{Data Availability}
The data that supports the findings of this study are openly available at Zenodo \cite{Zhu25data}.
\appendix
\section{Derivation of an envelope equation for the linear electrostatic ITG eigenmode}
\label{sec:app}
Here, we follow the WKB theory of Dewar and Glasser \cite{Dewar83} and present a derivation of the envelope equation in $\alpha$ for the linear electrostatic ITG eigenmodes. The general idea goes as follows. The linear collisionless gyrokinetic equation \eref{eq:GTC_linear} can be written as 
\begin{equation}
    \hat{\mc{L}}(\omega,\mc{E},\mu,\sigma,\bd{R},\nabla)\delta\!f=0,
\end{equation}
where $\hat{\mc{L}}$ is a linear operator and we have replaced $\pd_t$ with $-\rmi\omega$. Let us write $\delta\!f=\delta\!\hat{f}\rme^{\rmi S/\rho_*}$, so that
\begin{equation}
\label{eq:appA_general}
    \hat{\mc{L}}\delta\!f=(\hat{\mc{L}}_0\delta\!\hat{f}+\rho_*\hat{\mc{L}}_1\delta\!\hat{f})\rme^{\rmi S/\rho_*}=0,
\end{equation}
where
\begin{equation}
    \hat{\mc{L}}_0=\hat{\mc{L}}(\nabla\to \rmi\bd{k}/\rho_*+\nabla_\parallel),\quad \bd{k}=\nabla S, 
\end{equation}
and $\hat{\mc{L}}_1$ is the remaining parts in $\hat{\mc{L}}$ that involve $\nabla_\perp$. Let $\delta\!\hat{f}=\delta\!\hat{f}_0+\rho_*\delta\!\hat{f}_1$, the lowest-order term in \eref{eq:appA_general} is
\begin{equation}
    \hat{\mc{L}}_0\delta\!\hat{f}_0=0, 
\end{equation}
which is a first-order differential equation along field lines that gives the local dispersion relation. The next-order terms in \eref{eq:appA_general} are
\begin{equation}
\label{eq:appA_general_order0}
    \hat{\mc{L}}_0 \delta\!\hat{f}_1+\hat{\mc{L}}_1\delta\!\hat{f}_0=0.
\end{equation}
We would like to get rid of $\delta\!\hat{f}_1$ and obtain an  equation that only involves $\delta\!\hat{f}_0$ (the ``solubility constraint''  \cite{Dewar83}). To do so, we look for an inner product $\avg{.,.}$ and an adjoint operator $\hat{\mc{L}}_0^\dagger$ such that
\begin{equation}
\label{eq:appA_general_adjoint}
\avg{f,\hat{\mc{L}}_0g}=\avg{g,\hat{\mc{L}}_0^\dagger f}. 
\end{equation}
If such inner product and adjoint operator exist, then we can look for a solution $\delta\!\hat{f}_0^\dagger$ which satisfies $\hat{\mc{L}}_0^\dagger\delta\!\hat{f}_0^\dagger=0$. Then, we apply the inner product of $\delta\!\hat{f}_0^\dagger$ with \eref{eq:appA_general_order0} and apply the relation \eref{eq:appA_general_adjoint}, obtaining
\begin{equation}
\label{eq:appA_general_envelope}
\avg{\delta\!\hat{f}_0^\dagger,\hat{\mc{L}}_1\delta\!\hat{f}_0}=0.
\end{equation}
Equation \eref{eq:appA_general_envelope} is the envelope equation for $\delta\! \hat{f}_0$, which describes the mode structures across field lines. More discussions on the inner product and the adjoint operator for electromagnetic gyrokinetic systems can be found in \cite{Acton24}. 

In the following, we apply the above procedure to the collisionless electrostatic gyrokinetic equation with adiabatic electrons. We separate $\delta\! f$ into  adiabatic and nonadiabatic parts (assuming $Z_\rmi=1$):
\begin{equation}
    \delta\!f=-\bar{\varphi} f_0+h,
\end{equation}
where $\varphi=e\delta\Phi/T_\rmi$ is obtained from quasineutrality \eref{eq:GTC_poisson}:
\begin{equation}
\label{eq:appA_poisson}
n_\rmi(1+\tau)\varphi=\int \rmd\bd{v}\,\hat{J}_0 h,
\end{equation}
and $\bar{\varphi}=\hat{J}_0\varphi$. Here, $\tau=T_\rmi/T_\rme$, $\hat{J}_0$ is the gyroaverage operator, and $\rmd\bd{v}=\sum_{\sigma=\pm 1}(2\pi B/m_\rmi^2|v_\parallel|)\rmd\mu\rmd\mc{E}$.  Then, letting $\bd{h}=(h,\bar{\varphi})$, we can write   \eref{eq:GTC_linear} as
\begin{equation}
\label{eq:appA_GK}
\hat{\mc{L}}\bd{h}=(\rmi v_\parallel\pd_l+\omega-\hat{\omega}_\rmd)h-f_0(\omega-\hat{\omega}_*)\bar{\varphi}=0,
\end{equation}
where $\hat{\omega}_\rmd=-\rmi\bd{v}_\rmd\cdot\nabla$, $\hat{\omega}_*=-\rmi\bd{v}_*\cdot\nabla$, and 
\begin{equation}
    \bd{v}_*=\frac{T_\rmi}{ eBf_0}\frac{\pd f_0}{\pd\psi}(\hat{\bd{b}}\times\nabla\psi).
\end{equation}
Write $\bd{h}=(\bd{h}_0+\rho_*\bd{h}_1)\rme^{\rmi S/\rho_*}$ with $\bd{h}_0=(\hat{h}_0,\hat{\varphi}_0)$ and  $\bd{h}_1=(\hat{h}_1,\hat{\varphi}_1)$, the corresponding local gyrokinetic equation is 
\begin{equation}
\label{eq:appA_L0}
    \hat{\mc{L}}_0\bd{h}_0=(\rmi v_\parallel\pd_l+\omega-{\omega}_\rmd)\hat{h}_0-f_0(\omega-{\omega}_*)\hat{\varphi}_0=0,
\end{equation}
where $\omega_\rmd=\bd{k}\cdot\bd{v}_\rmd/\rho_*$ and $\omega_*=\bd{k}\cdot\bd{v}_*/\rho_*$. In particular, $\omega_*$ only depends on $k_\alpha$ and $\mc{E}$:
\begin{equation}
    \omega_*=\frac{k_\alpha T_\rmi}{\rho_* e}\left[\frac{\pd\ln n_\rmi}{\pd\psi}+\frac{\pd\ln T_\rmi}{\pd\psi}\left(\frac{\mc{E}}{T_\rmi}-\frac{3}{2}\right)\right].
\end{equation}
The next-order term $\hat{\mc{L}}_1$ are the parts in $\hat{\mc{L}}$ that linearly involve the $\nabla_\perp$ operator:
\begin{equation}
\label{eq:appA_L1}
\rho_*\hat{\mc{L}}_1\bd{h}_0=-\hat{\omega}_\rmd\hat{h}_0+f_0\hat{\omega}_*\hat{\varphi}_0.
\end{equation}
Let us define the inner product as
\begin{equation}
    \avg{f,g}=\int\rmd l\,\rmd\bd{v} \frac{fg}{f_0(\omega-\omega_*)},
\end{equation}
and define $\bd{f}=(f,\varphi_f)$ and $\bd{g}=(g,\varphi_g)$, so that
\begin{equation}
\label{eq:appA_inner_product}
\eqalign{
\avg{f,\hat{\mc{L}}_0\bd{g}}
\\
=\int\rmd l\,\rmd\bd{v}\left[\frac{\rmi v_\parallel f\pd_l g+(\omega-\omega_\rmd)fg}{f_0(\omega-\omega_*)}-f\varphi_g\right]
\\
=\int\rmd l\,\rmd\bd{v}\left[\frac{-\rmi v_\parallel g\pd_l f+(\omega-\omega_\rmd)gf}{f_0(\omega-\omega_*)}-g\varphi_f\right]
\\
=\avg{g,\hat{\mc{L}}_0^\dagger \bd{f}}.
}
\end{equation}
Here, we have applied an integration by part in $l$ for the term $\rmi v_\parallel f\pd_l g$, which can be done because $v_\parallel$ is canceled by the factor $1/|v_\parallel|$ in $\rmd\bd{v}$ and both $f_0\propto\exp(-\mc{E}/T_\rmi)$ and $\omega_*$ do no depend on $l$. The boundary terms in $\l$ are assumed to vanish, which is justified for passing particles if perturbations vanish in the good-curvature regions, and for trapped particles where the boundary terms cancel each other. In deriving \eref{eq:appA_inner_product} we also used the relation
\begin{equation}
    \int\rmd\bd{v} f\varphi_g=\int\rmd\bd{v} g\varphi_f,
\end{equation}
which can be easily seen from \eref{eq:appA_poisson}.
Therefore, we find the adjoint operator and its solution to be
\begin{equation}
\hat{\mc{L}}_0^\dagger=\hat{\mc{L}}_0(v_\parallel\to-v_\parallel),\quad \hat{h}_0^\dagger=\hat{h}_0(v_\parallel\to-v_\parallel).
\end{equation}
Finally, the envelope equation is $\avg{\hat{h}_0^\dagger,\hat{\mc{L}}_1\bd{h}_0}=0$, which from \eref{eq:appA_L1} can be written as
\begin{equation}
\label{eq:appA_general_envelope_equation}
    \int\rmd l\,\rmd\bd{v}\frac{\rmi\hat{h}_0^\dagger}{(\omega-\omega_*)}\left[\frac{\bd{v}_\rmd\cdot\nabla\hat{h}_0}{f_0}-\bd{v}_*\cdot\nabla\hat{\varphi}_0\right]=0.
\end{equation}
Equation \eref{eq:appA_general_envelope_equation} is an integro-differential equation and depends quadratically on $\hat{h}_0$. In order to proceed and obtain further analytic insights, certain assumptions are needed to simplify this equation.

In the following, we simplify the above derivations by considering the fluid limit $|\omega|\gg|v_\parallel\pd_l|$ and compare the results with \sref{sec:theory}. In the fluid limit we ignore  $v_\parallel\pd_l$  in $\hat{\mc{L}}$. Then, from  \eref{eq:appA_L0}, we obtain
\begin{equation}
\label{eq:appA_local_solution}
    \hat{h}_0=f_0\frac{\omega-\omega_*}{\omega-\omega_\rmd}\hat{\varphi}_0,
\end{equation}
which together with \eref{eq:appA_poisson} yields the local dispersion relation
\begin{equation}
\label{eq:appA_local_dispersion}    n_\rmi(1+\tau)=\int\rmd\bd{v}f_0J_0^2\frac{\omega-\omega_*}{\omega-\omega_\rmd}.
\end{equation}
To carry out the velocity-space integration, we use normalized velocity $\bd{x}=\bd{v}/v_{\rmt\rmi}$, so that $J_0=J_0(x_\perp\sqrt{2b})$ with $b=|k_\alpha a\nabla\alpha|^2/2$, $\omega_\rmd=k_\alpha\omega_{\rmd 0}(x_\perp^2/2+x_\parallel^2)/\rho_*$, and $\omega_*=k_\alpha\omega_{*0}[1+\eta(x^2-3/2)]/\rho_*$ with $\eta=\pd\ln T_\rmi/\pd\ln n_\rmi$. We also assume a large temperature gradient, $\omega_*\gg\omega_\rmd$, so that $\omega\sim\sqrt{\omega_*\omega_\rmd}$ and
\begin{equation}
    \frac{\omega-\omega_*}{\omega-\omega_\rmd}\approx-\frac{\omega_*}{\omega}+1-\frac{\omega_*\omega_\rmd}{\omega^2}.
\end{equation}
Also, in the long-wavelength limit, $b\ll 1$, we have
\begin{equation}
    J_0^2(x_\perp\sqrt{2b})\approx 1-bx_\perp^2.
\end{equation}
The integration can be carried out with the  following relations:
\begin{equation}
\int\frac{\rmd\bd{x}\,\rme^{-x^2}}{\pi^{3/2}}
\left(
\eqalign{
1\\
x_\perp^2\\
x^2-\frac{3}{2}\\
x_\perp^2(x^2-\frac{3}{2})\\
\frac{x_\perp^2}{2}+x_\parallel^2\\
(\frac{x_\perp^2}{2}+x_\parallel^2)(x^2-\frac{3}{2})\\
x_\perp^2(\frac{x_\perp^2}{2}+x_\parallel^2)\\
x_\perp^2(\frac{x_\perp^2}{2}+x_\parallel^2)(x^3-\frac{3}{2})
}
\right)
=
\left(
\eqalign{
1\\
1\\
0\\
1\\
1\\
1\\
\frac{3}{2}\\
3
}
\right),
\end{equation}
obtaining \eref{eq:theory_dispersion} from the simple model:
\begin{equation}
\label{eq:appA_local_dispersion_simple}
    \tau-\frac{bk_\alpha\omega_{T0}}{\rho_*\omega}+\frac{k_\alpha^2\omega_{\rmd 0}\omega_{T0}}{\rho_*^2\omega^2}\approx 0.
\end{equation}
Here, $\omega_{T0}=\eta\omega_{*0}$, and we assumed zero density gradient so that $\omega_{*0}\to0$ but $\eta\to\infty$ and $\omega_{T0}$ is still finite. To the next order in $\rho_*$ we still have
\begin{equation}
    \hat{\mc{L}}_1\bd{h}_0+\hat{\mc{L}}_0\bd{h}_1=0,
\end{equation}
Since the $\pd_l$ term has been ignored for $\hat{\mc{L}}_0$, we can obtain the envelope equation in a much simpler way:
\begin{equation}
\int\rmd\bd{v}\frac{J_0}{\omega-\omega_d}\hat{\mc{L}}_1\bd{h}_0=-\int\rmd\bd{v}\frac{J_0}{\omega-\omega_d}\hat{\mc{L}}_0\bd{h}_1=0,
\end{equation}
where the second equality comes from \eref{eq:appA_poisson} and \eref{eq:appA_local_dispersion}. Using \eref{eq:appA_local_solution}, this leads to
\begin{equation}
\label{eq:appA_local_envelope}
  \int\frac{\rmd\bd{v}f_0J_0}{\omega-\omega_\rmd}\left[{v}_\rmd^\alpha\pd_\alpha\left(\frac{\omega-\omega_*}{\omega-\omega_\rmd}\hat{\varphi}_0\right)-{v}_*^\alpha\pd_\alpha\hat{\varphi}_0\right]=0,  
\end{equation}
where $\bd{v}_\rmd^\alpha=\bd{v}_\rmd\cdot\nabla\alpha=\omega_{\rmd 0}(x_\perp^2/2+x_\parallel^2)$ and $\bd{v}_*^\alpha=\bd{v}_*\cdot\nabla\alpha=\omega_{*0}[1+\eta(x^2-3/2)]$. After carrying out the velocity-space integral, \eref{eq:appA_local_envelope} becomes an linear differential equation in $\alpha$ that describes the structure of $\hat{\varphi_0}$ across field lines. To compare with the results in \sref{sec:theory}, we assume $\omega_*\gg\omega\gg\omega_\rmd$, $\eta\gg1$, and $b=0$ so that $J_0=1$. At $\mc{O}(\eta)$, \eref{eq:appA_local_envelope} becomes 
\begin{equation}
\label{eq:appA_local_envelope2}
    -\frac{\omega_{T0}}{\omega}\left[\pd_\alpha(k_\alpha\hat{\varphi}_0)+k_\alpha\pd_\alpha\hat{\varphi}_0\right]=0,
\end{equation}
which leads to $\pd_\alpha\ln\hat{\varphi}_0=-\pd_\alpha \ln k_\alpha/2$, consistent with \eref{eq:theory_envelope} at $g=0$.  We can also assume $b\neq 0$ but $\omega_{\rmd0}=0$, in which case \eref{eq:appA_local_envelope} results in
\begin{equation}
\label{eq:appA_local_envelope3}
    \frac{\omega_{T0}}{\omega}\left(b\pd_\alpha\hat{\varphi}_0+\frac{1}{2}\hat{\varphi}_0\pd_\alpha b\right)=0.
\end{equation}
From \eref{eq:appA_local_dispersion_simple} we have $b\propto k_\alpha^{-1}$, and hence \eref{eq:appA_local_envelope3} leads to $\pd_\alpha\ln\hat{\varphi}_0=\pd_\alpha \ln k_\alpha/2$, which is somewhat different from \eref{eq:theory_envelope} at $f=0$. 
\section*{References}
\bibliographystyle{iopart-num}
\bibliography{references}

\providecommand{\newblock}{}
\begin{thebibliography}{10}
\expandafter\ifx\csname url\endcsname\relax
  \def\url#1{{\tt #1}}\fi
\expandafter\ifx\csname urlprefix\endcsname\relax\def\urlprefix{URL }\fi
\providecommand{\eprint}[2][]{\url{#2}}

\bibitem{Beurskens21}
Beurskens M, Bozhenkov S~A, Ford O, Xanthopoulos P, Zocco A, Turkin Y, Alonso
  A, Beidler C, Calvo I, Carralero D {\em et~al.\/} 2021 {\em Nuclear Fusion\/}
  {\bf 61} 116072

\bibitem{Beer95}
Beer M~A, Cowley S and Hammett G 1995 {\em Physics of Plasmas\/} {\bf 2} 2687

\bibitem{Highcock12}
Highcock E 2012 {\em The zero-turbulence manifold in fusion plasmas\/} Ph.D.
  thesis University of Oxford

\bibitem{Martin18}
Martin M~F, Landreman M, Xanthopoulos P, Mandell N~R and Dorland W 2018 {\em
  Plasma Physics and Controlled Fusion\/} {\bf 60} 095008

\bibitem{Dewar98}
Dewar R~L and Hudson S~R 1998 {\em Physica D: Nonlinear Phenomena\/} {\bf 112}
  275

\bibitem{Smoniewski21}
Smoniewski J, S{\'a}nchez E, Calvo I, Pueschel M~J and Talmadge J~N 2021 {\em
  Physics of Plasmas\/} {\bf 28} 042503

\bibitem{Sanchez21}
S{\'a}nchez E, Garc{\'\i}a-Rega{\~n}a J~M, Ba{\~n}{\'o}n~Navarro A, Proll
  J~H~E, Moreno C~M, Gonz{\'a}lez-Jerez A, Calvo I, Kleiber R, Riemann J,
  Smoniewski J and Others 2021 {\em Nuclear Fusion\/} {\bf 61} 116074

\bibitem{Xanthopoulos14}
Xanthopoulos P, Mynick H, Helander P, Turkin Y, Plunk G, Jenko F, G{\"o}rler T,
  Told D, Bird T and Proll J 2014 {\em Physical Review Letters\/} {\bf 113}
  155001

\bibitem{Wilms23}
Wilms F, Navarro A~B and Jenko F 2023 {\em Nuclear Fusion\/} {\bf 63} 086004

\bibitem{Wilms24}
Wilms F, Navarro A~B, Windisch T, Bozhenkov S, Warmer F, Fuchert G, Ford O,
  Zhang D, Stange T, Jenko F {\em et~al.\/} 2024 {\em Nuclear Fusion\/} {\bf
  64} 096040

\bibitem{Kornilov04}
Kornilov V, Kleiber R, Hatzky R, Villard L and Jost G 2004 {\em Physics of
  Plasmas\/} {\bf 11} 3196

\bibitem{Sanchez19}
S{\'a}nchez E, Estrada T, Velasco J, Calvo I, Cappa A, Alonso A,
  Garc{\'\i}a-Rega{\~n}a J, Kleiber R, Riemann J {\em et~al.\/} 2019 {\em
  Nuclear Fusion\/} {\bf 59} 076029

\bibitem{Cole19}
Cole M~D~J, Hager R, Moritaka T, Dominski J, Kleiber R, Ku S, Lazerson S,
  Riemann J and Chang C~S 2019 {\em Physics of Plasmas\/} {\bf 26} 082501

\bibitem{Wang20}
Wang H~Y, Holod I, Lin Z, Bao J, Fu J~Y, Liu P~F, Nicolau J~H, Spong D and Xiao
  Y 2020 {\em Physics of Plasmas\/} {\bf 27} 082305

\bibitem{BanonNavarro20}
Ba{\~n}{\'o}n~Navarro A, Merlo G, Plunk G, Xanthopoulos P, Von~Stechow A,
  Di~Siena A, Maurer M, Hindenlang F, Wilms F and Jenko F 2020 {\em Plasma
  Physics and Controlled Fusion\/} {\bf 62} 105005

\bibitem{Sanchez23}
S{\'a}nchez E, Ba{\~n}{\'o}n~Navarro A, Wilms F, Borchardt M, Kleiber R and
  Jenko F 2023 {\em Nuclear Fusion\/} {\bf 63} 046013

\bibitem{Jost01}
Jost G, Tran T, Cooper W, Villard L and Appert K 2001 {\em Physics of
  Plasmas\/} {\bf 8} 3321

\bibitem{Cole20}
Cole M~D~J, Moritaka T, Hager R, Dominski J, Ku S and Chang C~S 2020 {\em
  Physics of Plasmas\/} {\bf 27} 044501

\bibitem{Chen25a}
Chen H, Wei X, Zhu H and Lin Z 2025 {\em Nuclear Fusion\/} {\bf 65} 074002

\bibitem{Chen25b}
Chen H, Wei X, Zhu H and Lin Z {\em in preparation\/}

\bibitem{Nicolau25}
Nicolau J~H, Wei X, Liu P, Choi G and Lin Z 2025 {\em Nuclear Fusion\/} {\bf
  65} 086049

\bibitem{Xanthopoulos16}
Xanthopoulos P, Plunk G, Zocco A and Helander P 2016 {\em Physical Review X\/}
  {\bf 6} 021033

\bibitem{Helander15}
Helander P, Bird T, Jenko F, Kleiber R, Plunk G, Proll J, Riemann J and
  Xanthopoulos P 2015 {\em Nuclear Fusion\/} {\bf 55} 053030

\bibitem{Landreman22}
Landreman M and Paul E 2022 {\em Physical Review Letters\/} {\bf 128} 035001

\bibitem{Gonzalez22}
Gonz{\'a}lez-Jerez A, Xanthopoulos P, Garc{\'\i}a-Rega{\~n}a J, Calvo I,
  Alcus{\'o}n J, Navarro A~B~N, Barnes M, Parra F and Geiger J 2022 {\em
  Journal of Plasma Physics\/} {\bf 88} 905880310

\bibitem{Zocco16}
Zocco A, Plunk G, Xanthopoulos P and Helander P 2016 {\em Physics of Plasmas\/}
  {\bf 23} 082516

\bibitem{Zocco20}
Zocco A, Plunk G and Xanthopoulos P 2020 {\em Physics of Plasmas\/} {\bf 27}
  022507

\bibitem{Dewar83}
Dewar R and Glasser A 1983 {\em The Physics of Fluids\/} {\bf 26} 3038

\bibitem{Barnes19}
Barnes M, Parra F~I and Landreman M 2019 {\em Journal of Computational
  Physics\/} {\bf 391} 365

\bibitem{Mandell24}
Mandell N~R, Dorland W, Abel I, Gaur R, Kim P, Martin M and Qian T 2024 {\em
  Journal of Plasma Physics\/} {\bf 90} 905900402

\bibitem{Fu21}
Fu J~Y, Nicolau J~H, Liu P~F, Wei X~S, Xiao Y and Lin Z 2021 {\em Physics of
  Plasmas\/} {\bf 28} 062309

\bibitem{Nicolau21}
Nicolau J~H, Choi G, Fu J, Liu P, Wei X and Lin Z 2021 {\em Nuclear Fusion\/}
  {\bf 61} 126041

\bibitem{Singh23}
Singh T, Nicolau J~H, Nespoli F, Motojima G, Lin Z, Sen A, Sharma S and Kuley A
  2023 {\em Nuclear Fusion\/} {\bf 64} 016007

\bibitem{Zhu25}
Zhu H, Lin Z and Bhattacharjee A 2025 {\em Journal of Plasma Physics\/} {\bf
  91} E28

\bibitem{Lee87}
Lee W~W 1987 {\em Journal of Computational Physics\/} {\bf 72} 243

\bibitem{Romanelli89}
Romanelli F 1989 {\em Physics of Fluids B: Plasma Physics\/} {\bf 1} 1018

\bibitem{Biglari89}
Biglari H, Diamond P and Rosenbluth M 1989 {\em Physics of Fluids B Plasma
  Physics\/} {\bf 1} 109

\bibitem{Plunk14}
Plunk G, Helander P, Xanthopoulos P and Connor J 2014 {\em Physics of
  Plasmas\/} {\bf 21} 032112

\bibitem{Rodriguez25}
Rodr\'iguez E and Zocco A 2025 {\em Journal of Plasma Physics\/} {\bf 91} E21

\bibitem{Dewar97}
Dewar R 1997 {\em Plasma Physics and Controlled Fusion\/} {\bf 39} 453

\bibitem{Kotschenreuther95}
Kotschenreuther M, Rewoldt G and Tang W 1995 {\em Computer Physics
  Communications\/} {\bf 88} 128

\bibitem{Gaur23}
Gaur R, Abel I~G, Dickinson D and Dorland W~D 2023 {\em Journal of Plasma
  Physics\/} {\bf 89} 905890112

\bibitem{Gaur24}
Gaur R, Conlin R, Dickinson D, Parisi J~F, Dudt D, Panici D, Kim P, Unalmis K,
  Dorland W~D and Kolemen E 2024 {Omnigenous stellarator equilibria with
  enhanced stability, arXiv:2410.04576}

\bibitem{Zhu25data}
Zhu H 2025 {Data for the paper ``Global linear drift-wave eigenmode structures
  on flux surfaces in stellarators: ion temperature gradient mode'' [Data set].
  Zenodo. https://doi.org/10.5281/zenodo.15564134}

\bibitem{Acton24}
Acton G, Barnes M, Newton S and Thienpondt H 2024 {\em Journal of Plasma
  Physics\/} {\bf 90} 905900406

\end{thebibliography}
\end{document}